\documentclass[5p,twocolumn,11pt]{elsarticle}
\usepackage{amsbsy}
\usepackage{amssymb}
\usepackage{amsmath}
\usepackage{graphicx}
\usepackage{natbib}
\usepackage{hyperref}
\usepackage{tikz}
\usepackage{acronym}
\usepackage{pgfplots}
\usepackage{color}
\usepackage{xcolor}
\pgfplotsset{compat=1.5}

\newcommand{\be}{\begin{equation}}
\newcommand{\ee}{\end{equation}}
\newcommand{\bear}{\begin{eqnarray}}
\newcommand{\eear}{\end{eqnarray}}
\newcommand{\derpar}[1] {\frac{\partial}{\partial #1}}

\definecolor{niceblue}{RGB}{57,106,177}
\definecolor{nicered}{RGB}{204,37,41}

\acrodef{MKL}[MKL]{Math Kernel Library}

\journal{Computer Physics Communications}
\begin{document}

\begin{frontmatter}

\title{Magneto-thermal evolution of neutron stars with coupled Ohmic, Hall and ambipolar effects via accurate finite-volume simulations}

\author[label1,label2,label4]{Daniele Vigan\`o}
\author[label1,label2,label3]{Alberto Garcia-Garcia}
\author[label3]{Jos\'e A. Pons}
\author[label1,label2]{Clara Dehman}
\author[label1,label2]{Vanessa Graber}

\address[label1]{Institute of Space Sciences (IECC-CSIC), Campus UAB, Carrer de Can Magrans s/n, 08193, Barcelona, Spain}
\address[label2]{Institut d'Estudis Espacials de Catalunya (IEEC), Carrer Gran Capit\`a 2--4, 08034 Barcelona, Spain}
\address[label4]{Institute of Applied Computing \& Community Code (IAC3), University of the Balearic Islands, Palma, 07122, Spain}
\address[label3]{Departament de F\'{\i}sica Aplicada, Universitat d'Alacant, Ap. Correus 99, 03080 Alacant, Spain}

\begin{abstract}
Simulating the long-term evolution of temperature and magnetic fields in neutron stars is a major effort in astrophysics, having significant impact in several topics. A detailed evolutionary model requires, at the same time, the numerical solution of the heat diffusion equation, the use of appropriate numerical methods to control non-linear terms in the induction equation, and the local calculation of realistic microphysics coefficients.
Here we present the latest extension of the magneto-thermal 2D code in which we have coupled the crustal evolution to the core evolution, including ambipolar diffusion. It has also gained in modularity, accuracy, and efficiency. We revise the most suitable numerical methods to accurately simulate magnetar-like magnetic fields, reproducing the Hall-driven magnetic discontinuities.
From the point of view of computational performance, most of the load falls on the calculation of microphysics coefficients.
To a lesser extent,  the thermal evolution part is also computationally expensive because it requires large matrix inversions due to the use of an implicit method.
We show two representative case studies: (i) a non-trivial multipolar configuration confined to the crust, displaying long-lived small-scale structures and discontinuities; and (ii) a preliminary study of ambipolar diffusion in normal matter. The latter acts on timescales that are too long to have relevant effects on the timescales of interest but sets the stage for future works where superfluid and superconductivity need to be included.
\end{abstract}

\begin{keyword}
MHD; neutron stars; magnetic fields; high resolution shock capturing
\end{keyword}
\end{frontmatter}

\newpage

\section{Introduction}

Neutron stars, the compact endpoints of massive stars, are born very hot and fast-rotating. In the most magnetized cases, the so-called magnetars, their gigantic magnetic energy powers most of the electromagnetic emission. Magnetic fields are at the origin of several effects:  i) they regulate the loss of their huge rotational energy via electromagnetic torque; ii) their dissipation provides a source of heat, via Joule effect, that keeps the surface temperature high and enhances the X-ray thermal emission; iii)  the evolution causes magnetic stresses, triggering instabilities which give rise to transient multi-wavelength phenomena. Thus, understanding the magnetic field dynamics in detail is of utmost importance for this class of sources.

The magneto-thermal evolution of neutron stars (see a recent review \cite{pons19}) relies on two evolution equations: the heat diffusion equation (at the base of the so-called cooling models, reviewed in \cite{potekhin15b}) and the induction equation. They are coupled and need a detailed specification of the local microphysics (neutrino emissivity, heat capacity, thermal and electrical conductivity) and the structure of the star, usually assumed as fixed throughout the neutron star's life.

The seminal papers in the Nineties describing and estimating the main effects of magnetism at play in magnetars \citep{goldreich92,thompson93,thompson96} laid the foundations for more quantitative studies. In the last 15 years, neutron star models dedicated to thermal evolution have been gradually incorporating the effects of magnetic fields, and numerical simulations have been increasing their complexity. These modeling efforts can be broadly separated into two types: focused on the crust, or on the core. The magnetic evolution in the solid crust is relatively easy to describe by assuming the ions to be fixed in their equilibrium positions in the solid lattice (i.e., neglecting its elasticity/plasticity). Under this approximation, the Maxwell  equations reduce to the {\it electron magnetohydrodynamics} (eMHD) limit \cite{huba91,huba03}, in which electrons are the only charged component free to move. Such equations apply also to other scenarios in plasma physics and astrophysics \citep{witalis86,deng01,mozer02,bard18,kunz04,pandey08,bethune16}.

The first eMHD simulations in axial symmetry \cite{hollerbach02,kojima12} were soon extended to include fixed stratification (i.e., radial dependence in the simplified electron density and electrical conductivity profiles) \cite{hollerbach04,gourgouliatos13,gourgouliatos14a,gourgouliatos14b,gourgouliatos15}. In parallel, other models included from the beginning a realistic stellar structure and consistently calculated microphysical inputs \cite{pons07b}. This paved the way to the first simulations with the magnetic evolution fully coupled to the cooling models, initially neglecting the Hall effect \cite{aguilera08a,aguilera08b,pons09}. A significant step forward was the incorporation of relativistic corrections and the combined effects of Ohmic dissipation and the Hall effect \cite{vigano12a}, still in axial symmetry. The latter, which is improved in this work, presents so far the only available magneto-thermal evolution
code with fully realistic microphysics.
Meanwhile, the first simulations of the magnetic evolution in 3D \cite{wood15,gourgouliatos16,gourgouliatos18,gourgouliatos20a} adapted the geo-dynamo code {\tt PARODY} \cite{dormy98} to the neutron star scenario, again with a fixed stellar structure and simplified microphysical coefficients, which are assumed having only a radial dependence. Other semi-analytical studies included modeling of the crustal plasticity in the eMHD equations  \cite{lander19,kojima20}. Direct applications to astrophysical scenarios stem from all these works (e.g.,  \cite{vigano12b,vigano13,geppert13,geppert14,gourgouliatos20b}).

In the core, the situation is more complex due to its multi-component nature and fundamental open issues remain about the formulation of the problem itself. Several studies have suggested that ambipolar diffusion \cite{goldreich92, shalybkov95} could be the driving mechanism behind field evolution in young magnetars \cite{thompson93, thompson96}, typically relying on estimates of the relevant timescales \cite{glampedakis11b, gusakov17, kantor18, ofengeim18}. Numerical analyses have been restricted to 1D \cite{hoyos08, hoyos10} and 2D \cite{castillo17, passamonti17a, bransgrove18, castillo20} so far but lack a consistent treatment of the thermal and magnetic evolution based on realistic microphysics.

In this paper we start closing this gap and provide new results on ambipolar diffusion in normal-matter neutron star cores, following the formalism by \cite{passamonti17b}. This will form the basis for future work that incorporates the presence of quantum condensates that are likely present in the interiors of mature neutron stars, but complicate the field evolution further and are beyond the scope of this paper \cite{graber15, elfritz16, passamonti17a, gusakov20}.

The aforementioned numerical works exploit the spherical symmetry of the background stellar structure by using spherical coordinates combined with finite-volume/finite-difference methods along the radial direction, where gradients of physical quantities are usually steep. In these studies, two main families of numerical methods have been used, according to the discretization of the induction equation in the angular direction. The most common approach relies on the spectral decomposition of potential functions in spherical harmonics; however, it requires an analytical manipulation of the equations \cite{geppert91}. The second family stems from \cite{vigano12a} and applies finite-volume methods to evolve the magnetic field components, allowing them to resolve the magnetic discontinuities. A third option existing in the literature \cite{bransgrove18} is represented by finite-difference simulations with a scalar potential formalism.

Besides the complex coupling between local microphysics, heat diffusion, and global magnetic evolution, the main challenge of fully consistent magneto-thermal simulations lies in the non-linearity of the induction equation. Here we aim at providing a detailed assessment of numerical ingredients helping build a stable finite-volume code able to simulate typical magnetars' conditions. We also describe for the first time the structure of the code and highlight the computational cost and scalability of its different components. The simulations are in axial symmetry and use a modular code structure which improves in accuracy and efficiency the different versions used by our group during the last 15 years.

The paper is structured as follows. In \S~\ref{sec:model} we briefly summarize the problem and the equations. In \S~\ref{sec:methods} we present the relevant numerical methods and ingredients. In \S~\ref{sec:computational} we carry out an analysis of such methods from a numerical point of view, a computational analysis of our code, and a performance study of the most important parts. Finally, we show two representative simulations in \S~\ref{sec:casestudies}; the first one focusing on crustal field evolution for a non-trivial initial topology and the second one dedicated to an analysis of ambipolar diffusion in the core. We draw conclusions and state our future works in \S~\ref{sec:conclusions}.

\section{Magneto-thermal models}\label{sec:model}

\begin{figure}[t]
	\centering
	\includegraphics[width=0.42\textwidth]{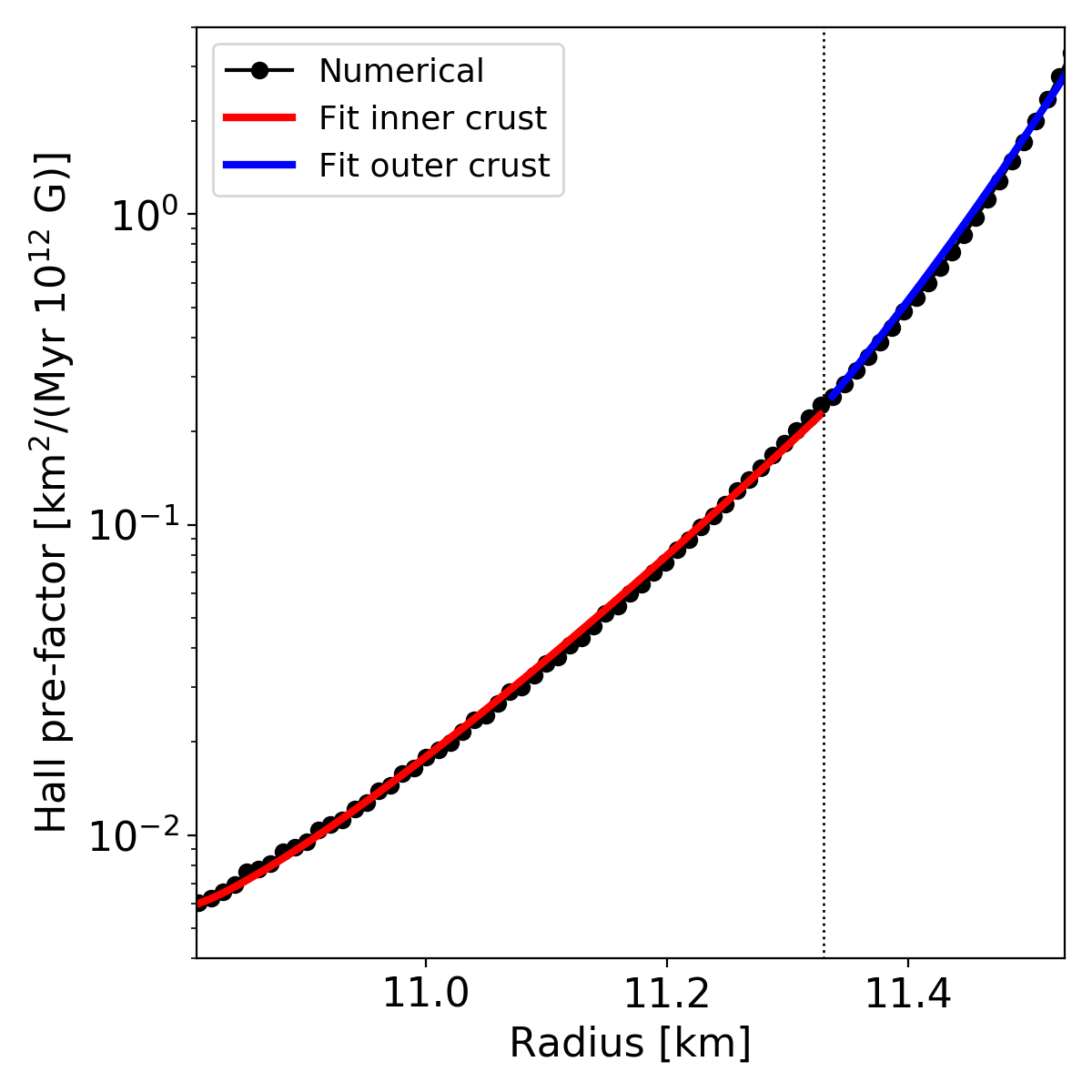}
	\caption{Hall prefactor $f_h$ for a realistic neutron star crust, considered up to a mass density $\rho=10^{10}$ g~cm$^{-3}$ (in this case corresponding to a radius $R_{10}=11.56$ km, using the SLy4 equation of state with $M=1.4~M_\odot$). The colored lines represent the fit with a piecewise function $f_{\rm h,fit}=f_0 \exp{[k(r-R_{\rm cc})^b]}$, where $R_{\rm cc}=10.81$ km is the crust-core interface and the parameters are: $f_0=0.011$, $k=10$, $b=1.8$ in the outer crust $11.33 \; {\rm km} <r<R_{10}$ (blue), and $f_0=0.006$, $k=8$, $b=1.2$ in the inner crust $R_{\rm cc}< r < 11.33$ km (red). Note that the specific fit depends on the equation of state and mass, but in general the functional form is super-exponential in radius, steepening in the outermost layers due to the decrease in density.
	}
	\label{fig:fh}
\end{figure}

\subsection{Background star's structure}

Realistic magneto-thermal models need to assume a background structure for the star in order to calculate necessary microphysical ingredients, such as the electron density $n_e$. The structure is provided by the Tolman-Oppenheimer-Volkoff equations \cite{oppenheimer39}, which solve the hydrostatic equilibrium assuming a static interior Schwarzschild metric $ds^2 = -e^{2\nu(r)} c^2dt^2 + e^{2\lambda(r)}(d\theta^2 + \sin^2\theta d\phi^2)$, where $e^{2\lambda}=1-2Gm(r)/(c^2r)$ and $\nu(r)$ is determined by $d\nu/dP=-(P(r)+\rho(r)c^2)^{-1}$, where $m(r)$ is the enclosed gravitational mass, $\rho$ is the energy density and $P$ is the pressure. The relativistic length correction $e^{\lambda}$ is hereafter included in the definition of the line and surface elements of the integrals and in the operators $\vec{\nabla}$ containing the radial derivatives (see e.g. \cite{pons19} for details). Note that the deviations from a spherically symmetric hydrostatic profile due to the inferred/observed values of magnetic fields and rotation are negligible for our purposes (see \cite{gomes19,haskell08} for magnetic deformations). For the crustal field evolution, an important quantity (as we will describe later) is the Hall prefactor $f_h=c/(4\pi e n_e)$. To contrast with the profile assumed by other works \cite{cumming04, gourgouliatos13, gourgouliatos15, gourgouliatos16}, in Fig.~\ref{fig:fh} we show the radial profile of $f_h$ for a typical star ($M=1.4~M_\odot$, SLy4 equation of state  \cite{douchin01}) employed in our simulations. It exhibits a super-exponential rise of about three orders of magnitude from the crust-core to the crust-envelope interface, here assumed to be at around $10^{10}$ g~cm$^{-3}$.

\subsection{Heat diffusion equation}

The heat diffusion equation governs the evolution of the temperature $T$ (see e.g. \cite{potekhin15b}). Within a given volume $V$ enclosed by a surface $S$, the integral form reads:
\begin{eqnarray}
&& \int_V c_\mathrm{v}\,\frac{\partial (T \mathrm{e}^\nu)}{\partial t} ~dV
+ \oint_S (\mathrm{e}^{2\nu} \vec{F}\cdot\hat{n})~dS = \nonumber\\
&& = \int_V \mathrm{e}^{2\nu} \left(\frac{j^2}{\sigma_e} - \dot{\epsilon}_\nu\right)~dV~,
\label{eq:heat_diffusion}
\end{eqnarray}
where several microphysical ingredients evaluated in the local frame (and generally dependent on density, temperature and magnetic field) are needed: $c_\mathrm{v}$ is the specific heat; the heat flux density $\vec{F}$ is obtained by the Fick's law:
\begin{equation}
\vec{F} = - \mathrm{e}^{-\nu} \hat{\kappa}\cdot\vec{\nabla}(\mathrm{e}^\nu T)~,\label{eq:heat_flux}
\end{equation}
where $\hat{\kappa}$ is the anisotropic thermal conductivity tensor; the source term includes the rate per unit volume of Joule heating $j^2/\sigma_e$, where $\sigma_e$ is the electrical conductivity parallel to magnetic field lines (see below) and neutrino losses $\dot{\epsilon}_\nu$. The electrical currents $j$ are calculated, at each point of the star, according to their definition (see \S~\ref{sec:magnetic_evolution}). The microphysical ingredients entering in eq.~(\ref{eq:heat_diffusion}) and ~(\ref{eq:heat_flux}) are summarized in the next subsection.

\subsection{Microphysics}

A complete revision of the microphysics entering in the magneto-thermal models is given in \cite{potekhin15b}. For the sake of brevity, here we simply summarize the main microphysical inputs to be computed, considering the background structure and the local values of temperature and magnetic field:

\begin{itemize}

\item {\it Thermal and electrical conductivities.}
The microphysical processes that contribute to the transport properties strongly depend on temperature and density. In the core, conductivities are very high, which implies that the core is basically isothermal (except in the first few decades after birth), and the electrical resistivity is orders of magnitude smaller than in the crust, implying much longer Ohmic timescales. In the outer crust (relatively low density) the dominant process is electron-phonon scattering, while electron-impurity scattering becomes the most relevant process in the inner crust for temperatures low enough.
For weak magnetic fields, the conductivity is isotropic. On the contrary, for high magnetic fields, the anisotropy is significant and the thermal conductivity is represented by a tensor. Its components are calculated using the public code released by A. Potekhin. We refer the interested reader to the website\footnote{\rm http://www.ioffe.ru/astro/conduct/} for more details and a complete list of references. In this work, we employ the 2019 release, slightly modified to switch off quantizing effects in the crust to speed up the calculations. Under strong quantizing fields, the real conductivity as a function of density oscillates about the classical values, due to the gradual filling of Landau levels. More details about the formalism can be found in Section 2 of \cite{potekhin15b}.
These oscillations are more prominent at very low density, but for our purposes, and for our spatial resolutions, the few percent corrections that the quantized prescription provides do not justify the increase in required computational time by one order of magnitude (considering that microphysics calculations are the computational bottleneck, see below).

\item  {\it Specific heat.}
The bulk of the total heat capacity of the neutron star is given by matter in the core, where most of the mass is contained. The crustal specific heat has contributions from the ion lattice, the degenerate electron gas, and the neutron gas in the inner crust (see \cite{aguilera08b} and references therein for the models used here, and \cite{page2012} for a detailed discussion). If neutrons appearing beyond the neutron drip point are not superfluid, they control the specific heat in the inner crust, but their contributions are exponentially  suppressed when the temperature drops below the neutron superfluid critical temperature \cite{Levenfish94}. For a detailed computation of the crustal specific heat we use the
publicly available codes\footnote{http://www.ioffe.ru/astro/EIP/}, describing the equation of state for a strongly magnetized, fully ionized electron-ion plasma \cite{potekhin10}. We also refer to the recent reviews \cite{potekhin15b,pons19} for more details.

\item {\it Neutrino emissivity.}
Neutrino emission processes drive the cooling of the star during the first $\sim 10^5$ years (neutrino cooling era), after which the star is cold enough to hamper neutrino production such that the surface photon emission dominates (photon cooling era). We use the same formulae for neutrino processes as described in Table 1 of \cite{potekhin15b}, where a detailed list of references can be found.

\item {\it Superfluidity.} We implement superfluidity corrections to the previous quantities for neutrons (singlet state) in the inner crust, and for neutrons (triplet) and protons (singlet) in the core. The critical temperature and the energy gap as a function of the Fermi momenta are approximated by the effective parametrization of \cite{kaminker01}, with different possible choices for the parameters, given by Table II of \cite{ho15}: we will show results for their models {\tt SFB, TTpa, CCDKp.}

\end{itemize}

\subsection{Magnetic field evolution equations}\label{sec:magnetic_evolution}

The integral form of Faraday's induction law for a surface $S$ reads (in Gaussian units):
\begin{equation}
\frac{\partial}{\partial t}\int_S  (\vec{B}\cdot \hat{n}) ~dS + c \oint_{\partial S}  (\mathrm{e}^{\nu} \vec{E})\cdot{d\vec{l}} = 0~,
\label{eq:induction}
\end{equation}
where $\hat{n}$ is the normal to the surface and $d\vec{l}$ is the line element along the surface border $\partial S$. The definition of the electric field $\vec{E}$ generally includes the electric currents $\vec{j}$, defined by Amp\`ere's law in its conservative form as
\begin{equation}\label{eq:current_mhd}
\int_S (\vec{j}\cdot\hat{n})~dS = \frac{c~\mathrm{e}^{-\nu} }{4 \pi} \oint_{\partial S} (\mathrm{e}^{\nu} \vec{B}) \cdot{d\vec{l}}~.
\end{equation}
Note that in axial symmetry, the poloidal-toroidal decomposition of any solenoidal field is particularly easy, $\vec{B}_{\rm tor} = B_\varphi \hat{\varphi}$ and $\vec{B}_{\rm pol} = B_r \hat{r} + B_\theta\hat{\theta}$, and each corresponding component of the currents depends only on the other magnetic field component: $\vec{j}=\vec{j}_{\rm pol}(\vec{B}_{\rm tor}) + \vec{j}_{\rm tor}(\vec{B}_{\rm pol})$.

\subsubsection{Crust}

In the crust, we include the Ohmic and the non-linear Hall term in the definition of the electric field:
\begin{equation}\label{eq:hall_induction}
ce^\nu \vec{E} = \eta \vec{\nabla}\times (\mathrm{e}^{\nu}\vec{B} )
+ f_h \left[ \vec{\nabla}\times (\mathrm{e}^{\nu}\vec{B}) \right]
\times \vec{B}~.
\end{equation}
The pre-factors on the right-hand side are the diffusivity $\eta=c^2/(4\pi \sigma_e)$ and the Hall prefactor $f_h$. Due to their inverse dependences on the electron density and electrical conductivity $\sigma_e$ respectively, they both vary by orders of magnitude across the crust, as shown in Fig.~\ref{fig:fh} for $f_H$.
The diffusivity is similarly steep in the radial direction and, since it includes temperature-dependent processes, it may also present angular variations up to a factor of a few. More importantly, $\sigma_e$ in the outer crust increases by orders of magnitude as the star cools down (see e.g. \cite{potekhin15b,pons19}). At the same time $B$ tends to decay, therefore the relative weight of the two terms in the electric field varies non-trivially in time. The associated timescales vary accordingly by orders of magnitude.

Note that our crustal induction equation neglects terms such as the thermo-electric effect \cite{geppert91}, relevant possibly only at high temperatures and in the outermost layers of the star (envelope).

\subsubsection{Core}\label{sec:core}

The core physics is more complex and having a consistent magnetic field evolution framework is not trivial, in particular due to the presence of superconducting protons. In this work, we include the ambipolar diffusion in normal-conducting, non-superfluid matter based on the formalism \cite{passamonti17b}. Ambipolar diffusion, a direct result of the core's multi-component nature, is caused by the relative motion between the charged particles and the neutrons. It can be incorporated into our field evolution model via an \textit{ambipolar velocity} $\vec{v}_{\rm a}$ that enters a generalized Ohm's law:
\begin{equation}\label{eq:Efield_ambipolar}
ce^\nu \vec{E} = \eta \vec{\nabla}\times (\mathrm{e}^{\nu}\vec{B} )
- \mathrm{e}^{\nu}\vec{v}_{\rm a} \times \vec{B}~,
\end{equation}
where $\vec{v}_{\rm a} \equiv x_{\rm n} (\vec{v}_{\rm p} - \vec{v}_{\rm n})$. Here, $x_{\rm n}$ denotes the neutron fraction and $\vec{v}_{\rm p}, \vec{v}_{\rm n}$ the proton and neutron velocities, respectively. We neglect the Hall term because strong coupling between the electrons and the protons renders it basically irrelevant in the core. However, we retain the ambipolar term, which looks like an advective term, but is highly non-linear (approximately cubic) in $B$, since the relative velocity between the charged components (protons and electrons) and neutrons is roughly proportional to the Lorentz force, as we discuss next (see also \cite{goldreich92}). Assuming equilibrium (i.e., neglecting the time derivatives in the momentum equations for each species), the ambipolar velocity can be defined by
\begin{equation}\label{eq:force_balance_ambipolar}
\vec{v}_{\rm a} = \frac{x_{\rm n}^2 \tau_{\rm pn}}{m_{\rm p}^*}
		\left[ \frac{\vec{f}_{\rm L}}{n_e} - \nabla (\Delta \mu)\right]~,
\end{equation}
where $\tau_{\rm pn}\propto T^{-2}\rho^{1/3}$ \cite{yakovlev90} represents the relaxation time for proton-neutron collisions, $m_{\rm p}^*$ the proton effective mass and $n_{\rm e}$ taken as the charge density (under the assumption of charge neutrality). While the Lorentz force $\vec{f}_{\rm L} \equiv (\vec{j} \times \vec{B})/c$ is straight forward to calculate, the chemical potential term requires an additional constraint. Only early on in a neutron star's life are $\beta$-reaction sufficiently fast to achieve equilibrium on dynamical timescales and thus $\Delta \mu \sim 0$. As we aim to model the magneto-thermal evolution over longer timescales, we require an additional equation for the chemical deviation. Following \cite{passamonti17a}, we solve the elliptic equation
\begin{equation}\label{eq:elliptic}
\nabla^2 (\Delta \mu) - \frac{1}{b} \frac{\partial \Delta \mu}{\partial r}
			- \frac{1}{a^2} \Delta \mu
		= \nabla \cdot \left( \frac{\vec{f}_{\rm L}}{n_e} \right)
			- \frac{1}{b} \frac{f^r_{\rm L}}{n_e}~,
\end{equation}
where the parameters $a$ and $b$, controlled by nuclear reaction rates and microphysics, have the dimension of length, and $f^r_{\rm L}$ is the radial component of the Lorentz force. They are defined as follows: $1/a^2:=\lambda_a \beta/x_n$, $1/b:=\frac{d\beta}{d r}$, where $\beta:=m_p^*/(x_n n_e \tau_{pn})$ and $\lambda_a=\lambda_a (T, \rho)$ is the coefficient describing the net $\beta$-decay reaction rate, linearized: $\Delta \Gamma \simeq \lambda_a \Delta \mu$. The $\lambda_a$ coefficient is $\propto T^4\rho^{1/3}$ for direct Urca processes or $\propto T^6\rho^{2/3}$ for modified Urca (see formulae (18)-(19) in \cite{passamonti17b} and references within). For a given, uniform temperature and magnetic field configuration, \cite{passamonti17a} calculated the corresponding velocity fields, imposing $v_a^r=0$ at the crust core interface. Since chemical imbalances can balance only the irrotational part of $\vec{f}_{\rm L}$, the velocity-field patterns are mostly irrotational when the star is hot ($T\simeq 10^9$ K, i.e., newly born stars), or solenoidal at temperatures of a few $10^8$ K, relevant for observed, middle-aged neutron stars.

We go beyond this analysis, solving the equations above to derive the ambipolar velocity at each numerical timestep (i.e., considering the evolving temperature to evaluate $\lambda_a$ and $\tau_{pn}$ at each point), and including it consistently into the core induction equation. This improves on recent works \cite{castillo17, castillo20}, by having more realistic microphysical coefficients, including the thermal evolution, and smoothly coupling the core evolution to the crust. As a first step, in this work we neglect the neutron velocity and the effects of superconductivity. The background neutron velocity has been taken into account by \cite{castillo17,castillo20}, who found a faster evolution in their specific models. Similarly, superconductivity makes $\tau_{pn}$ much longer than in the case considered here, thus giving much higher values of $\vec{v}_a$ (with, consequently, further computational challenges).

\subsection{Initial conditions}

The temperature at birth is $\sim 10^{11}-10^{12}$ K. Its precise value is not important for our purposes of long-term evolution, since the intense neutrino losses at such temperatures lead to the convergence to the same cooling curve after few years/decades.

For the magnetic field, instead, the initial condition is crucial: the first centuries will be dominated by transient whistler and Hall-drift waves if the solution is particularly out of eMHD equilibrium (which is the case also if we start from a MHD equilibrium) and if the initial configuration does not match smoothly with the chosen boundary conditions.

The initial magnetic field in our code is prescribed by using the scalar functions $\Phi$ and $\Psi$ for the poloidal and toroidal components as in \cite{geppert91,pons07b,pons09}, which easily allow for the definition of multipoles. The first model, {\tt Core}, includes the ambipolar diffusion and uses the same initial twisted-torus model as in \cite{passamonti17b}, where the toroidal field is contained within the closed field lines and, automatically, the azimuthal component of the Lorentz force is initially zero everywhere. In order to have faster dynamics and test the numerical methods needed for the Hall term, we also consider two models where the magnetic field is confined to the crust. In order to test the methods under different topologies, we consider two cases, shown in Table~\ref{tab:models}: one, {\tt CrP}, dominated by a poloidal dipole and one, {\tt CrM}, consisting of a mix (with a similar weight) of the first three multipoles $l=1,2,3$, with similar content of energy between poloidal and toroidal components. The general dynamics of the {\tt CrP} case are well known, leading to an equatorial discontinuity on the $B_r$ and $B_\varphi$ components. Less trivial models, like {\tt CrM}, are much less studied and we will mainly consider that case in our assessment.

\begin{table}
	\begin{tabular}{l | c c c c c}
		\hline
		\hline
		& $B_{\rm dip}$ & $E_{\rm mag}$ & $\frac{E_{\rm mag}^{\rm tor}}{E_{\rm mag}}$ & Poloidal & Toroidal \\
		& [G] & [$10^{45}$ erg] & $\%$ & $l$ & \\
		\hline
		{\tt Core}   & $10^{14}$ & $5.2$ & 2.9 & $1$ & torus \\
		{\tt CrP}	& $10^{14}$ & $16$ & 0.6 & $1$ & $l=2$ \\
		{\tt CrM}	& $10^{13}$ & $1.9$ & 54  & $1,2,3$ & $l=1,2,3$ \\
		\hline
		\hline
	\end{tabular}
	\caption{Summary of the initial configurations of the models considered: dipolar poloidal field at the polar surface, total magnetic energy, toroidal magnetic energy fraction, non-zero multipoles in the poloidal component, and toroidal topology.}
	\label{tab:models}
\end{table}

Note that all the options considered in this paper and in all previous numerical simulations are arguably unrealistic (e.g., crust-confined fields, or large-scale, smooth dipole/quadrupole+twisted torus). As a matter of fact, the dynamo mechanism during and just after the collapse \cite{thompson93}, should lead to a much more complex topology, likely characterized by a repartition of the energy over a spectrum of multipoles both in the core and in the crust, no axial symmetry and possibly an off-set of the magnetic moment from the center. This is an open issue from both a theoretical and a numerical point of view.

\subsection{Boundary conditions}

In both the heat diffusion and induction equations, the interaction with the external environment plays an important role. The outermost layers host the steepest gradients in the structure and temperature profiles. Therefore, the timescales there are much shorter than in the interior: it is numerically unfeasible to directly evolve the magnetic field and temperature up to the star's surface. The usual approach, undertaken here as well, is to include the envelope as a boundary condition, implicitly assuming that, because of the much shorter thermal relaxations timescales, the temperature profiles very quickly adjust to the equilibrium solution.

Regarding the thermal evolution, we rely on hydrostatic envelope models with a given composition (light or heavy elements), obtained separately for a set of internal temperatures (at the bottom of the envelope) and magnetic fields. For a given composition and assuming an emission model (blackbody in our case), this allows us to infer the surface temperature and flux at each point of the surface for the underlying internal temperature and magnetic field. We make use of the analytical fit to such models, as given by \cite{potekhin15a}, where more details about the envelope models can be found. The envelope model is important in controlling the photon emissivity, which is the dominant cooling mechanism at late ages ($\gtrsim 10^5$ yr).

For the magnetic field, we assume potential solutions as an external boundary condition, meaning no electrical currents circulating in the envelope and across the surface. We enforce this condition via multipole expansion of the radial magnetic field at surface, as almost all studies assume (but see the effect of a magnetosphere threaded by currents in \cite{akgun18b}).

Internally, for the models including the core, we impose a similar potential solution in the central cell (meaning simply that no currents can circulate right in the center). The difference is that at the surface we use the branch of solutions regular at infinity (each multipole $l$ goes like $B \propto r^{-(l+2)}$), while in the center the one regular at vanishing radius $(B \propto r^{(l-1)})$.

Finally, we impose at the axis reflecting boundary conditions on both temperature and magnetic field, derived by the axial symmetry assumption.

\subsection{Crust-core interface}\label{sec:cc_interface}

If the magnetic field is confined to the crust, we impose zero tangential electric field components at the interface between the crust and the core. This means that the radial magnetic field is kept to zero all the time, while the tangential component of the magnetic field can evolve. This naturally creates a current sheet that allows a discontinuity between a non-magnetized core and a crust threaded by currents. Since $\eta$ is discontinuous across the interface, the fine details of the treatment of the supercurrents affect the local deposition of heat. In our grid, the current sheet flows along a three-radial-point layer, and for simplicity is not considered in the Joule heating.

In the more realistic case of a core-threading magnetic field, the situation is much more intricate.
The crust is made of a lattice of very heavy nuclei, while the core is composed of a liquid phase of uniform nuclear matter (neutrons, protons, and electrons). In principle, these very different conditions allow for the electric field entering the induction equation to present a discontinuity in the radial direction if one has a sharp crust-core interface. However, in reality, there is arguably a transition layer, the pasta phase \cite{oyamatsu93}, whose transport properties are very uncertain (see \cite{lopez21} for a review). As one goes deeper into the inner crust, nuclei lose their regular shape, which could result in a higher electron resistivity \cite{horowitz15} (but see \cite{nandi18}). Conversely, as density increases, nuclei dissolve as we approach the uniform nuclear matter phase, and we could expect that the microphysical properties become more similar to those of the core.

For practical purposes, and considering our limited knowledge of details of the transition, we prescribe a smooth matching of the electric field components. We define a transition region of $\sim {\cal O}(10^2)$ meters (10 numerical points) around the crust-core interface. Within the transition region, we redefine the electric field via a cubic interpolation of the values of $\vec{E}$ appearing at its two extremes in the radial direction (i.e., inner crust and outer core).  This interpolation ensures that the radial profile of the three electric components and their radial derivatives are continuous. It substantially improves the stability of the code, avoiding the occurrence of unstable discontinuities at the interface.

\section{Numerical methods and ingredients}
\label{sec:methods}

\begin{figure}
  \centering
  \resizebox{!}{0.9\textheight}{
    \begin{tikzpicture}[node distance=2cm]

\usetikzlibrary{shapes.geometric, arrows}

\tikzstyle{startstop} = [rectangle, rounded corners, minimum width=3cm, minimum height=1cm,text centered, draw=black, fill=orange!30]
\tikzstyle{io} = [trapezium, trapezium left angle=70, trapezium right angle=110, minimum width=3cm, text width=2.5cm, minimum height=1cm, text centered, draw=black, fill=gray!30]
\tikzstyle{process} = [rectangle, minimum width=3cm, minimum height=1cm, text centered, text width=3cm, draw=black, fill=orange!30]
\tikzstyle{decision} = [diamond, minimum width=3cm, minimum height=1cm, text centered, draw=black, fill=yellow!30]
\tikzstyle{arrow} = [thick,->,>=stealth]

\node (start) [startstop] {Start};
\node (in1) [io, below of=start] {Input and Initialization};
\node (pro1) [process, fill=red!30, below of=in1] {Adaptive cooling timestep $dt_c$};
\node (pro4) [process, fill=blue!30, below of=pro1] {Electrical/Thermal conductivities};
\node (pro5) [process, fill=blue!30, below of=pro4] {Heat Capacity};
\node (pro6) [process, fill=blue!30, below of=pro5] {Neutrino Emissivity};
\node (pro7) [process, fill=red!30,below of=pro6] {Boundary Envelope model};
\node (pro8) [process, fill=green!30, below of=pro7] {Ambipolar velocity};
\node (prob0) [process, fill=red!30, below of=pro8] {Current and Electric field};
\node (prob1) [process, fill=green!30, below of=prob0] {Compute Joule};
\node (prob2) [process, fill=green!30, below of=prob1] {Magnetic Analysis};
\node (prob3) [process, fill=green!30, below of=prob2] {(Magnetic Stresses)};
\node (prob4) [process, fill=green!30, below of=prob3] {Adaptive Magnetic $dt_b$};
\node (prob5) [process, fill=green!30, below of=prob4] {Update $\vec{B}$};
\node (prob6) [process, fill=green!30, below of=prob5] {Magnetic Boundary Conditions};
\node (prob7) [process, fill=green!30, below of=prob6] {(Spin Period Evolution)};
\node (decb) [decision, below of=prob7, yshift=-1cm] {$t_b < t_c + dt_c$};
\node (proub) [process, process, fill=green!30, right of=decb, xshift=3cm] {Update $t_b$};
\node (pro9) [process, fill=red!30, below of=decb, yshift=-1cm] {Update $Te^\nu$};
\node (dect) [decision, below of=pro9, yshift=-1cm] {$t < t_{final}$};
\node (prout) [process, fill=red!30,right of=dect, xshift=5cm] {Update $t$};
\node (stop) [startstop, below of=dect, yshift=-1cm] {Stop};

\node (in2) [io, right of=in1, xshift=-8 cm, yshift=-4 cm]{Initial $\vec{B}$};
\node (in3) [io, below of=in2] {Superfluid Model};
\node (in4) [io, below of=in3]
{Impurities profile};
\node (in5) [io, below of=in4] {Star's structure \& grid};
\node (in6) [io, below of=in5] {Equation of state};

\draw [arrow] (start) -- (in1);
\draw [arrow] (in1) -- (pro1);
\draw [arrow] (pro1) -- (pro4);
\draw [arrow] (pro4) -- (pro5);
\draw [arrow] (pro5) -- (pro6);
\draw [arrow] (pro6) -- (pro7);
\draw [arrow] (pro7) -- (pro8);
\draw [arrow] (pro8) -- (prob0);
\draw [arrow] (prob0) -- (prob1);
\draw [arrow] (prob1) -- (prob2);
\draw [arrow] (prob2) -- (prob3);
\draw [arrow] (prob3) -- (prob4);
\draw [arrow] (prob4) -- (prob5);
\draw [arrow] (prob5) -- (prob6);
\draw [arrow] (prob6) -- (prob7);
\draw [arrow] (prob7) -- (decb);
\draw [arrow] (decb) -- node[anchor=east] {yes} (pro9);
\draw [arrow] (pro9) -- (dect);
\draw [arrow] (dect) -- node[anchor=east] {yes} (stop);
\draw [arrow] (dect) -- node[anchor=south] {no} (prout);
\draw [arrow] (decb) -- node[anchor=south] {no} (proub);

\draw [arrow] (prout) |- (pro1);
\draw [arrow] (proub) |- (prob0);
\draw [arrow] (dect) -- (stop);

\draw [arrow] (in2) |- (in1);
\draw (in2) -- (in3);
\draw (in3) -- (in4);
\draw (in4) -- (in5);
\draw (in5) -- (in6);

\end{tikzpicture}
  }
  \caption{Flowchart of the magneto-thermal 2D evolution code. Thermal evolution blocks are highlighted in red, microphysics parts in blue, and magnetic evolution steps in green. In parenthesis there are the optional by-products of the calculations.}
  \label{fig:flowchart}
\end{figure}
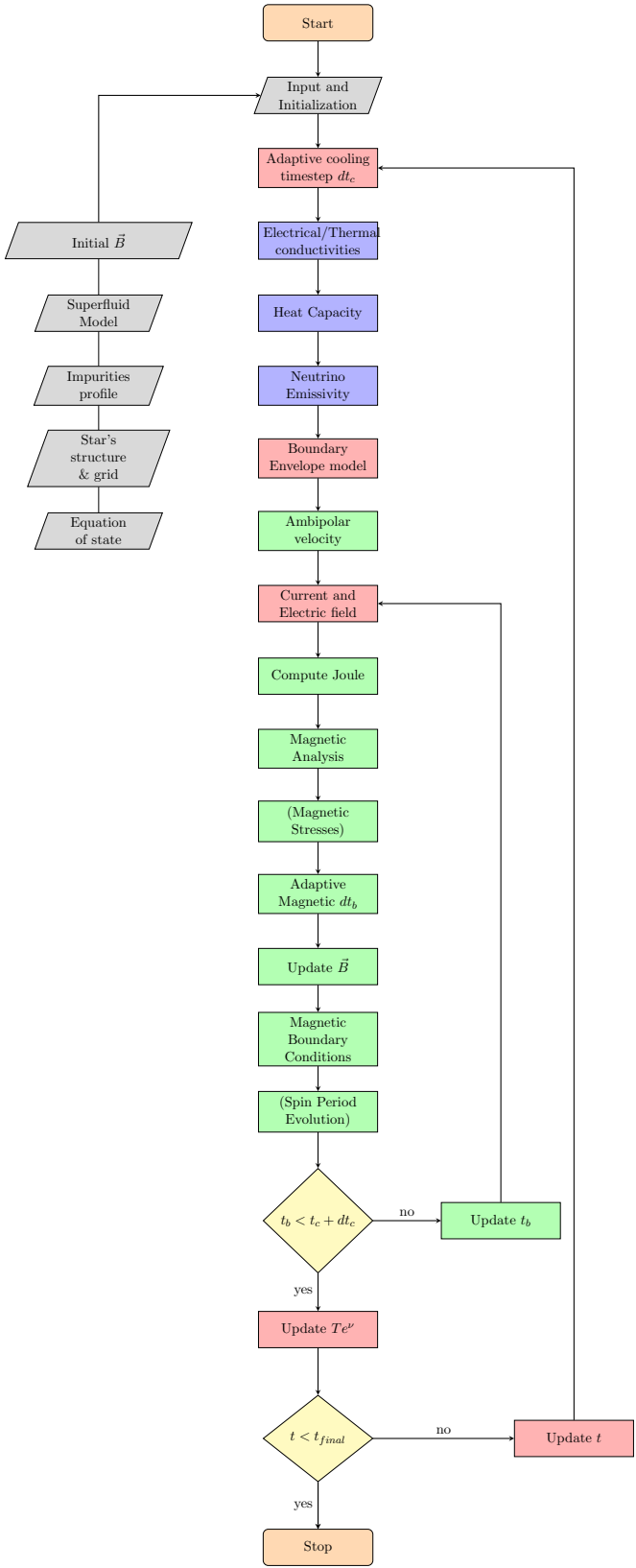

The magneto-thermal code is implemented in {\tt Fortran90/95} and represents an improvement of different versions of this code used previously by our group \cite{aguilera08a,pons07b,pons09,vigano12a,akgun18b}. The current iteration features a fully modular structure with an integrated {\tt{CMake}} build system that helps making the code easier to maintain, develop and extend. The codebase contains modules devoted to physics (thermal evolution, magnetic evolution and microphysics), others for data structures and support (grid and constants), utility ones (output) and an external module for making use of third-party libraries.

Figure \ref{fig:flowchart} shows a flowchart of the main program of the code. In the initialization, the star's structure is calculated for a given equation of state and central pressure, the input parameters are given (initial temperature, magnetic field strength and topology, impurity parameter, superfluid model, envelope model and numerical methods to be used), and some fixed quantities and mathematical functions (e.g., numerical grid, relativistic factors, geometrical elements and Legendre polynomials) are calculated. Then, the code enters into the main loop, within which there are three main parts: microphysical calculations, magnetic field evolution and thermal evolution. Note that the rotational evolution is a by-product coming from the evolution of the dipolar component at surface, $B_p(t)$. Therefore, it can also be performed as a post process, for a given spin-down formula including or not inclination angle evolution \cite{pons19}. Similarly, one can calculate as a by-product the magnetic stresses and, knowing the maximum shear from microphysics, estimate the frequency of crustal failure events, as in \cite{perna11,pons11,dehman20}.

The methods we describe here are based on the conservative formulation shown in \S~\ref{sec:model}, applying the Stokes' and Gauss theorems to each numerical cell, as described in detail in \cite{pons19}. Below, we go through a series of numerical ingredients that allow to increase the accuracy, numerical stability and the efficiency of the code.

\subsection{Discretization on the grid}

In axial symmetric problems involving a stably stratified star, the spherical coordinates $(r,\theta,\varphi)$ are the natural choice since they allow us to discretize the star in radial layers (see \ref{app:cartesian} for the issues arising from the implementation of the model in Cartesian coordinates). The cells cover the star from the center to the putative crust-envelope interface, called bottom of the envelope or, for simplicity, surface $R_\star$, where we apply the boundary conditions for both evolution equations. This interface, strictly speaking, moves outward in time, due to the gradual freezing of the outer layers, as locally the temperature drops below the melting value. However, for practical purposes, we consider a fixed grid and simulate the star down to densities $\rho_b \sim 10^{10}$ g~cm$^{-3}$. According to the cooling models, at such a density, the freezing happens at $T\sim 0.5-1\times 10^9$ K, corresponding typically to an age of a few decades: ideally, one would need to reach one to two orders of magnitude less in density, to cover the entire crust at middle ages ($\lesssim 10^5$ yr). However, the numerical timestep and stability constraints arising from the steep rise of $f_h$ and $\eta$ put limitations on the location of such an interface.

\begin{figure}[!t]
	\centering
	\includegraphics[width=\linewidth, trim={1.25cm 0 0 0}, clip]{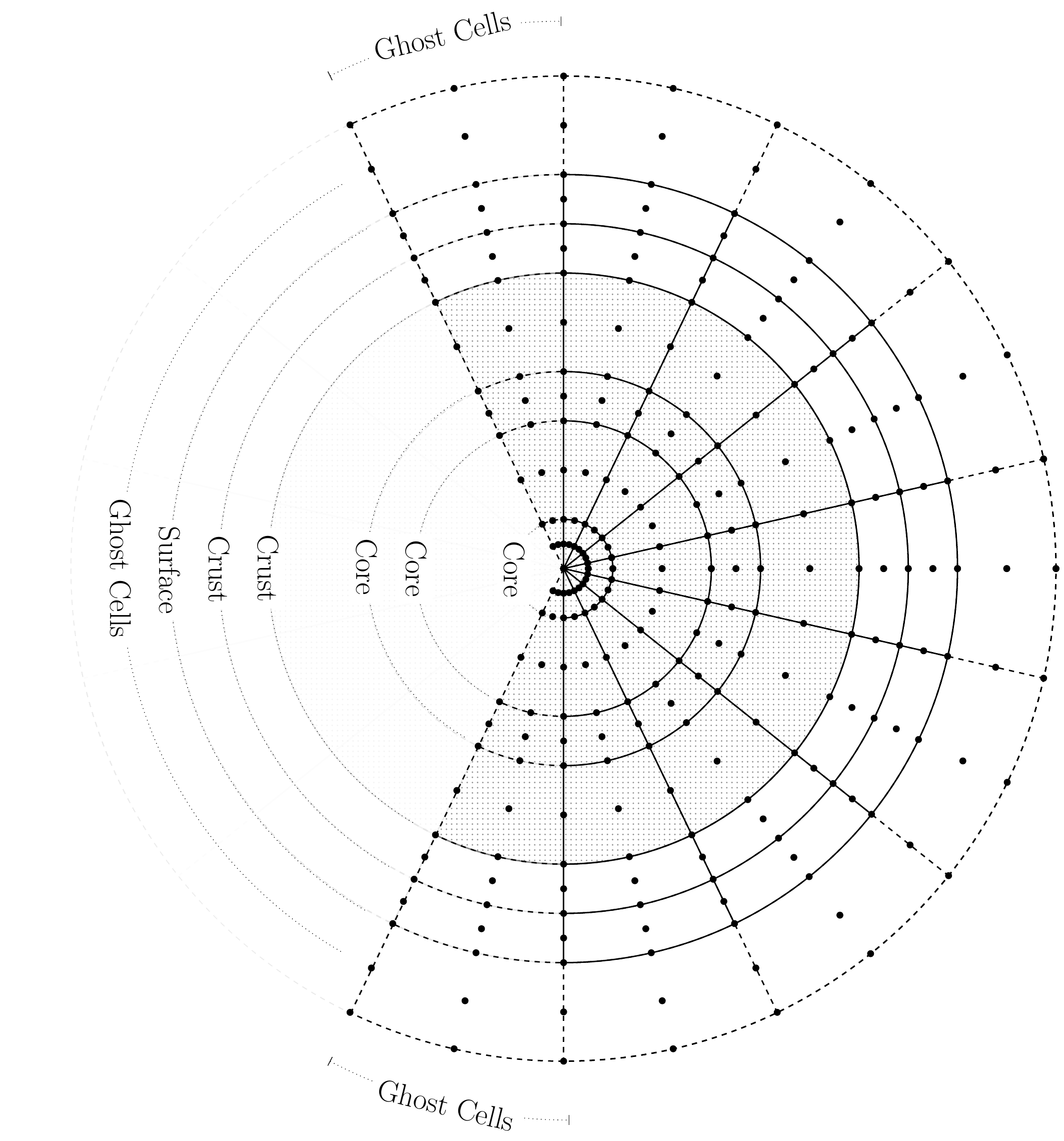}
	\caption{Numerical grid, represented through a simplified meridional cut of one hemisphere. The black points indicate all locations where the three components of the magnetic field, electric currents and electric fields are defined, while the temperature is only evolved at the center of the cells delimited by the solid lines. Ghost cells used in the code to impose boundary conditions are indicated with dashed lines.}
	\label{fig:grid}
\end{figure}

The radial size of the cells, $dr$, needs to be much finer in the crust than in the core, since in the outer layers the radial gradients of the background (density, pressure) and temperature profiles are much larger than in the inner ones. We define a smooth transition from a large step in the core to a small step for the crust by a function $dr(r) \propto  1 - 0.5\Delta_{dr}\tanh[(r - r_0)/(R_{\star}\sigma_t)]$ where $\Delta_{dr} \in [0,2)$ denotes the contrast (i.e., the relative difference between large and small steps), $r_0$ and $\sigma_t$ mark the position and size of the transition region respectively. In this paper, we set $r_0 = 0.8 R_\star$ (thus, well below the crust-core interface), $\Delta_{dr}=0.8$ and $\sigma_t=0.1$. We have made sure that results do not depend on these parameters, as long as a radial resolution $\lesssim 50$ m in the outer crust is granted.

The angular step size, $d\theta$, is instead taken as constant. Under axial symmetry, the axis is treated with standard reflective boundary conditions in the angular direction. In general, the angular gradients tend to be much smaller than the radial ones (especially in the crust), reason why $dr$ is chosen to be $\sim {\cal O}(10 {m})$ in the crust, while $r~d\theta \sim {\cal O}(100 {m})$.

We sketch the grid in Fig.~\ref{fig:grid}, indicating with lines the meridional section of the cells, which are $N_r$ and $N_\theta$ in the meridional and radial direction, respectively. Typically, we use $N_r=100$ (of which 37 lie in the crust) and $N_\theta=49$ (the latter being odd in order to have a cell centered at the equator). The temperature values that are evolved lie at the center of the cells, while the heat fluxes are defined in the middle of their interfaces. A simple average between the first neighbors is used when temperature values are needed on a cell's vertexes or interfaces (in order to evaluate the electrical/thermal conductivities or the temperature gradients appearing in the heat flux).

On the other hand, magnetic fields, electric fields and currents are defined and evolved at the black points in Fig.~\ref{fig:grid}: center, middle of the interfaces and vertexes of each cell. The points where the magnetic field is evolved are therefore $2N_r$ and $2N_\theta-1$ in the radial and angular direction, respectively. This contrasts with previous versions \cite{vigano12a}, which used one staggered grid where electric and magnetic components were defined in displaced locations, naturally arising from the discretized conservative form of the equation for a cell. The advantage of electro-magnetic fields evolved on a full grid is that no interpolations are needed, since all components are defined everywhere. However, the method still relies on Stokes' theorem applied to the interfaces centered on the evolved point.

At each point labeled by the angular and radial indexes $(i,j)$, we define the volume cell $V^{(i,j)}$, the interface areas normal to each $k$ direction, $S_k^{(i,j)}$, and the line elements along each $m$-direction $dl_m^{(i,j)}$. Such elements are widely used in the discretization version of eqs.~\ref{eq:heat_diffusion}, \ref{eq:induction} and \ref{eq:current_mhd} (see \cite{pons19} for the definitions).

Importantly, note that the full grid used, being effectively a superposition of two staggered grids in each direction, conserves exactly the divergence, like a standard staggered grid. As a matter of fact, the Gauss theorem applied to the evolution of $\vec{\nabla}\cdot\vec{B}$, together with the induction equation (\ref{eq:induction}), reads:

\begin{equation}
\frac{d(\vec{\nabla}\cdot\vec{B})^{(i,j)}}{dt}= \frac{c}{V^{(i,j)}} \sum_{{(k,l)=(i\pm 1,l),(i,j\pm 1)}}\oint_{\partial S^{(k,l)}}e^\nu E^m dl_m~,
\end{equation}
where the sum is performed over the six surfaces delimiting a given cell at ($i,j$), two of which (those having a normal in the azimuthal direction) do not give any net contribution due to axial symmetry. For each surface, we consider the line integral of the elements of circulation $E^m dl_m$. Such elements are located at one of the staggered points ($i,j\pm 1$, $i\pm 1,j$), so that each of them appears twice with opposite sign. Therefore, they cancel out and the right-hand side is zero by construction, exactly like in a standard staggered grid.

The fact that the full grid is effectively composed by double staggered grids in each dimension also means that the numerical results actually consist of the co-existence of two numerical solutions. The origin of the double solution is that the magnetic field on odd points is determined by the electric field on even points and vice versa, and the numerical boundaries of the odd and even points are slightly different by definition. The solutions are coupled only partially by the Hall term.
As a consequence, we see from our simulations that the results tend to show odd-even decoupling. This issue is substantially cured by: (i) imposing as a boundary condition a linear interpolation among the two radially-neighboring points at the point just below $R_\star$ (and above $R_{\rm cc}$ for crust-confined models), for the toroidal components of both the vector potential $A_\varphi$ (from which the poloidal field is calculated) and magnetic field $B_\varphi$; (ii) adding hyper-resistivity, especially at late stages (see below).

\subsection{Cooling scheme and microphysics}

The heat diffusion equation can be solved by standard methods for parabolic equations with stiff terms, since the neutrino emissivities are highly nonlinear with the temperature, $\propto T^\alpha$, with $\alpha \in [5,8]$ \cite{potekhin15b}. The Joule term $\propto \sigma_e(T)^{-1}$ can also be treated as stiff, even though the dependence with $T$ is less dramatic. Such stiffness is well managed by implicit methods relying on the linearization of the source term and the inversion of the tridiagonal block matrix ${\cal M}$, which relates the updated set of $N_r\times N_\theta$ values $\tilde{T}_a^{n+1}$, to the previous set $\tilde{T}_a^n$, where $\tilde{T}\equiv Te^\nu$ is the redshifted temperature and $a$ labels each cell: ${\cal M}^{ab}\tilde{T}_b^{n+1}=v_a(\tilde{T}^n)$, where $v_a$ is the vector which also collects the old temperatures, the sources and the dependencies of the sources on the local temperatures $T_a^n$. The elements of the matrix arise from the discretization of the problem on a spherical coordinate grid and the use of standard centered differences to evaluate the gradients in the heat flux $\vec{F}$.

After less than a century the core becomes isothermal (constant $\tilde{T}$, see e.g. Fig. 3 of \cite{potekhin18}). Thus, we solve the equations at all core points only until $\tilde{T}$ is homogeneous (relative differences less than $0.1\%$). After that (approximately at 100 yr), we instead consider the core as one radial layer only, which leads to a substantial computational time saving. We do that by considering the correct weighted average of the specific heat and neutrino emissivity in the core, but evolve only one temperature, considering the thermal conductivity only at the crust-core interface. We made sure that the results converge to the case where we evolve all points.

Temperatures are not allowed to be smaller than $10^6$ K, because the microphysics implemented are not suitable for such regimes. Therefore, $10^6$ K is taken as a floor value, which means that the cooling model can follow the star up to $\sim 10^6$ yr maximum.

\subsection{Adaptive timestep}

\begin{figure}[t]
	\centering
	\includegraphics[width=0.45\textwidth]{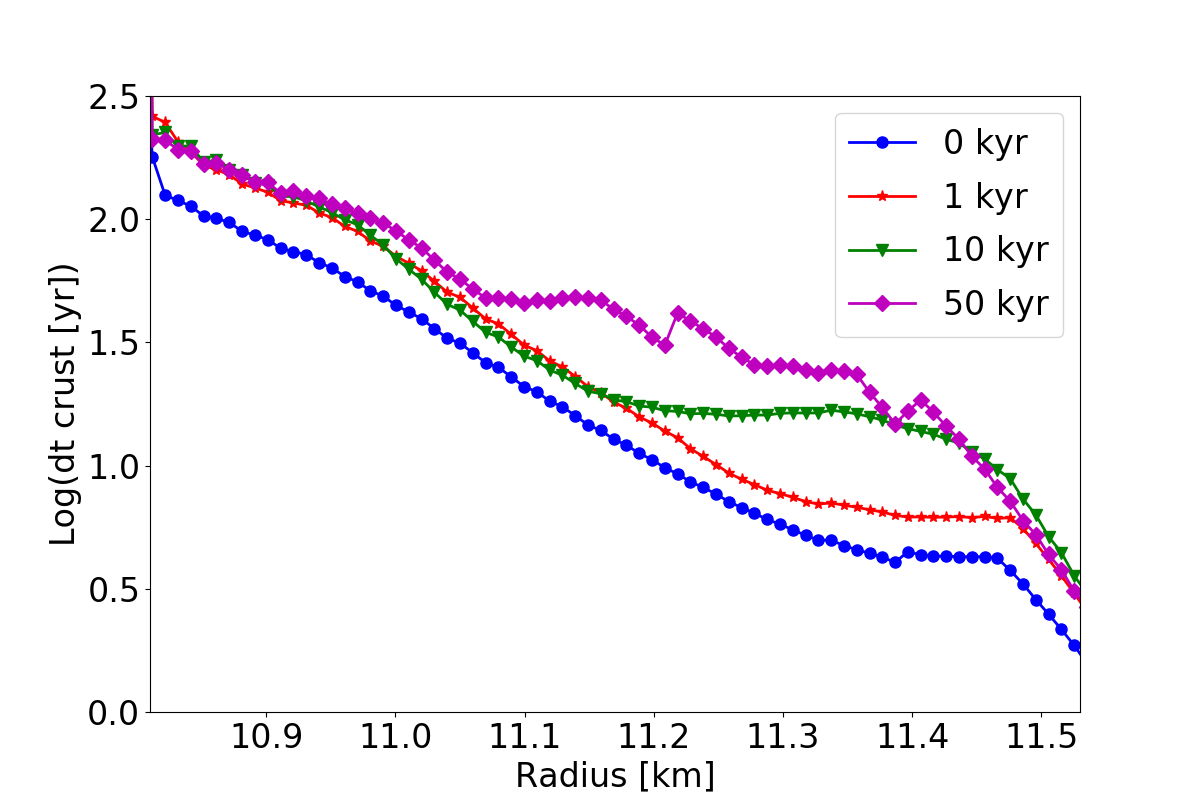}
	\caption{Typical evolution of $dt_{\rm crust}(r)$ (considering the minimum over $\theta$ at each $r$), for model {\tt crM}, at different times. Note that in general the minimum is located in the last cells.}
	\label{fig:dt_local}
\end{figure}

The cooling and magnetic timescales vary a lot during the star's life. As the neutron star cools down, there are two effects: on one side, neutrino emissivities drop by many orders of magnitude; on the other side, the matter becomes more thermally and electrically conductive. As a consequence, using a fixed timestep over Myr-long times would incur in an unnecessary large computational cost: it is advisable to adopt two different dynamical definitions of the numerical timestep, one for each equation.

For the cooling, one can use a timestep, $dt_c$, increasing with time, since the temperature variations are much larger at the beginning. In our case we use typically a phenomenologically increasing value, starting with $dt_c=10^{-2}$ yr during the first years (when the drop in temperature is very fast), increasing it to $dt_c \sim t/100$ until it reaches a large value which is kept uniform, $dt_c \sim 100-1000$ yr. This choice is not fine-tuned for optimization and could be more elegant, but it is a practical implementation that ensures stability in the implicit scheme described above.

More complicated is the timestep used in the magnetic evolution ($dt_b$) since the intensity and topology of the magnetic field define the Ohmic, Hall and ambipolar timescales, together with the conductivity and the electron density. Any precise assessment of the Courant-limited maximum value for $dt_b$ is hampered by the non-linearity of the problem. As a matter of fact, the characteristic velocities of the eMHD equations can be obtained only in their linearized version (see \cite{vigano19}), i.e., perturbations on top of a background field, which is not the case of our realistic scenario. Therefore, we simply introduce the local estimate on dimensional grounds in the crust and in the core, as:

\begin{eqnarray}
&&  dt_{\rm crust}(r,\theta) =  \frac{(\min\{dr,rd\theta\})^2}{\eta(r,\theta) + f_h(r) B(r,\theta)}~,\\
&&  dt_{\rm core}(r,\theta) =  \frac{\min\{dr,rd\theta\}}{|\vec{v}_a|}~.
\end{eqnarray}
The spatial resolution enters quadratically in the crustal estimate, in agreement with the non-linear dispersion relation of the whistler waves in eMHD. The timestep can then be defined dynamically at each step as follows:

\begin{equation}\label{eq:dtb_kc}
dt_b = k_{\rm cour} \min_{r,\theta}\{dt_{\rm crust}(r,\theta),dt_{\rm core}(r,\theta)\}~,
\end{equation}
where $k_{\rm cour}$ is a constant pre-factor that ideally depends only on the numerical scheme and needs to be tuned, as we will see below. The magnetic timestep is severely constrained by three factors: (i) high resolution, (ii) high magnetic field, (iii) low values of $dt_{\rm crust}$ of the outermost layers, where the denominator is systematically the largest. In Fig.~\ref{fig:dt_local} we show the evolution of $\min_{\theta} \{dt_{\rm crust}\}(r)$ for the model {\tt CrM}. Its value steeply decreases from centuries in the inner crust to fractions of years in the outermost layers, thus constraining the above-mentioned crust-envelope interface, $\rho_b$, to be $\sim 10^{10}$ g~cm$^{-3}$ at most, to make the computation feasible (see e.g. the discussion in \S2 of \cite{potekhin18}). mentioned above. In model {\tt Core}, $dt_{\rm core}$ is orders of magnitude larger than $dt_{\rm crust}$, which remains the timestep bottleneck (this would probably not hold anymore if superconductivity was accounted for in the ambipolar velocity).

For magnetar-like values of $B$, $dt_b$ has to be much smaller than $dt_c$ chosen above. Therefore, each cooling timestep embeds many magnetic timesteps, visible as the green nested loop in the flowchart Fig.~\ref{fig:flowchart}. The microphysical ingredients are updated together with the temperatures, so that the electrical conductivity appearing in the induction equation changes every $dt_c$, and not every $dt_b$ (the second option would incur in a notable additional computational cost with a limited gain in accuracy).

\subsection{Time advance schemes}

We have implemented and compared four different time advance methods:

\begin{itemize}
	\item Simple Euler (EUL), with which all components of $\vec{B}$ are advanced just by multiplying $dt_b$ with the increment of the magnetic field, $\delta \vec{B}$.
	\item Alternate Euler (EULA) as in \cite{vigano12a}, in which: (i) $\vec{B}_{\rm tor}$ is evolved from $\vec{E}_{\rm pol}$; (ii) the evolved $\vec{B}_{\rm tor}$ is used to calculate $\vec{j}_{\rm pol}$ and update $\vec{E}_{\rm tor}$, which now depends on a mix of old and updated components; (iii) $\vec{B}_{\rm pol}$ is evolved using the intermediate $\vec{E}_{\rm tor}$. This alternate advance actually corresponds to introducing an implicit hyper-resistive-like term (proportional to fourth-order derivatives) in the poloidal components of the induction equation \cite{toth08}.
	\item Fourth-order-accurate Runge-Kutta (RK4);
	\item Fourth-order-accurate Adams-Bashforth (AB4), which considers the combination of the increments $\delta\vec{B}$ of the current and the three previous timesteps. The first three timesteps at the beginning of the simulation are evolved by EUL method (this choice does not really affect the results or the stability, being restrained to three steps only).
\end{itemize}
We will assess the optimal performance of each method for a given set-up and initial conditions, based on the maximum value of $k_{\rm cour}$ we can set without having numerical instabilities or loss of convergence.

Note that the numerical errors are always dominated by the space discretization, unless one is able to keep very close to the maximum Courant time (which is impossible in our realistic, complex scenario). Therefore, the accuracy of the solution does not depend on the time advance method, which instead shows different performance in terms of stability (see \S~\ref{sec:computational}). As in other contexts, such differences arise from the fact that each time discretization method can implicitly add some numerical diffusivity which stabilizes the solution.

\subsection{Toroidal magnetic field advance}

We now consider the spatial discretization of the induction equation. First, we consider two options for the time advance of the toroidal magnetic field, which in axial symmetry coincides with the azimuthal component, $\vec{B}_{\rm tor} = B_\varphi~\hat{\varphi}$:

\begin{itemize}
	\item[a.] The use of the poloidal electric field $\vec{E}_{\rm pol}$ within the simplest discretization of Eq.~(\ref{eq:induction}):
\begin{eqnarray}
&& \frac{\partial B_\varphi^{(i,j)}}{\partial t}\frac{S_\varphi^{(i,j)}}{c} = \nonumber \\
&& = (e^\nu E_r dl_r)^{(i+1,j)}  - (e^\nu E_r dl_r)^{(i-1,j)}  \nonumber \\
&& +  (e^\nu E_\theta dl_\theta)^{(i,j-1)} - (e^\nu E_\theta dl_\theta)^{(i,j+1)}~.
\label{eq:bphi_advance_electric}
\end{eqnarray}
where the quantities in parentheses are the elements of the electric field circuitation and are evaluated at $(i\pm 1,j)$ and $(i,j\pm 1)$, i.e., the first neighboring cells in the angular and radial direction.

\item[b.] The use of a finite difference for the part of the electric field containing the Hall term, known for its Burgers-like behavior in axial symmetry \citep{vainshtein00,vigano12a}. In this case the toroidal component of the induction Eq (\ref{eq:induction}) can be re-written as:
\item[]
\begin{eqnarray}\label{eq:induction_burgers}
&& \frac{\partial B_\varphi}{\partial t} + \frac{\lambda_r}{f_h e^\lambda}\derpar{r}\left(\frac{f_h e^\nu B_\varphi^2}{2}\right) + \\
&& + \frac{\lambda_\theta}{r}\derpar{\theta}\left(\frac{B_\varphi^2}{2}\right) + \frac{c}{S_{\varphi}} \oint_{\partial S_{\varphi}}  (\mathrm{e}^{\nu} \vec{E}_{\rm res})\cdot{d\vec{l}} = 0 ~, \nonumber
\end{eqnarray}

where $\vec{E}_{\rm res}= \vec{j}_{\rm pol}/\sigma_e$ is the resistive part of the poloidal electric field, its circuitation is discretized as in Eq.~(\ref{eq:bphi_advance_electric}), and we have defined

\begin{eqnarray}
&& \lambda_r = - 2f_h \frac{\cot\theta}{r}~, \label{eq:induction_burgers_lambda1}\\
&& \lambda_\theta = - r^2\frac{e^{2\nu}}{e^\lambda}\derpar{r}\left( \frac{f_h}{e^\nu r^2}\right)~.\label{eq:induction_burgers_lambda2}
\end{eqnarray}

In analogy with Burgers' equation, the factors $f_h e^\nu B_\varphi^2/2$ (or simply $B_\varphi^2/2$ for the $\theta$-direction) can be interpreted as the flux, and $\lambda_rB_\varphi$ and $\lambda_\theta B_\varphi$ have velocity dimension. Note that the solenoidal constraint is still maintained thanks to axial symmetry: $\vec{B}_{\rm tor}(r,\theta)$ does not contribute to the divergence.
\end{itemize}

\begin{figure}[!htb]
  \centering
  \includegraphics[width=0.45\textwidth]{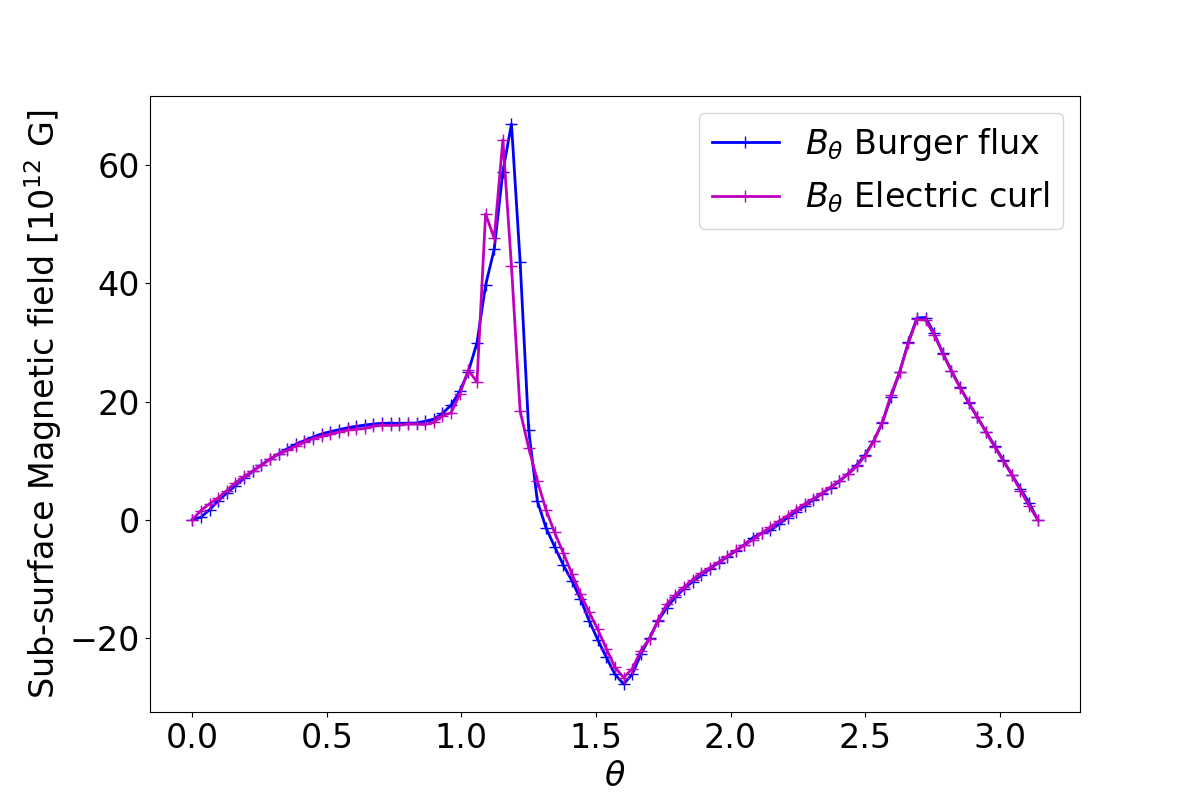}
  \includegraphics[width=0.45\textwidth]{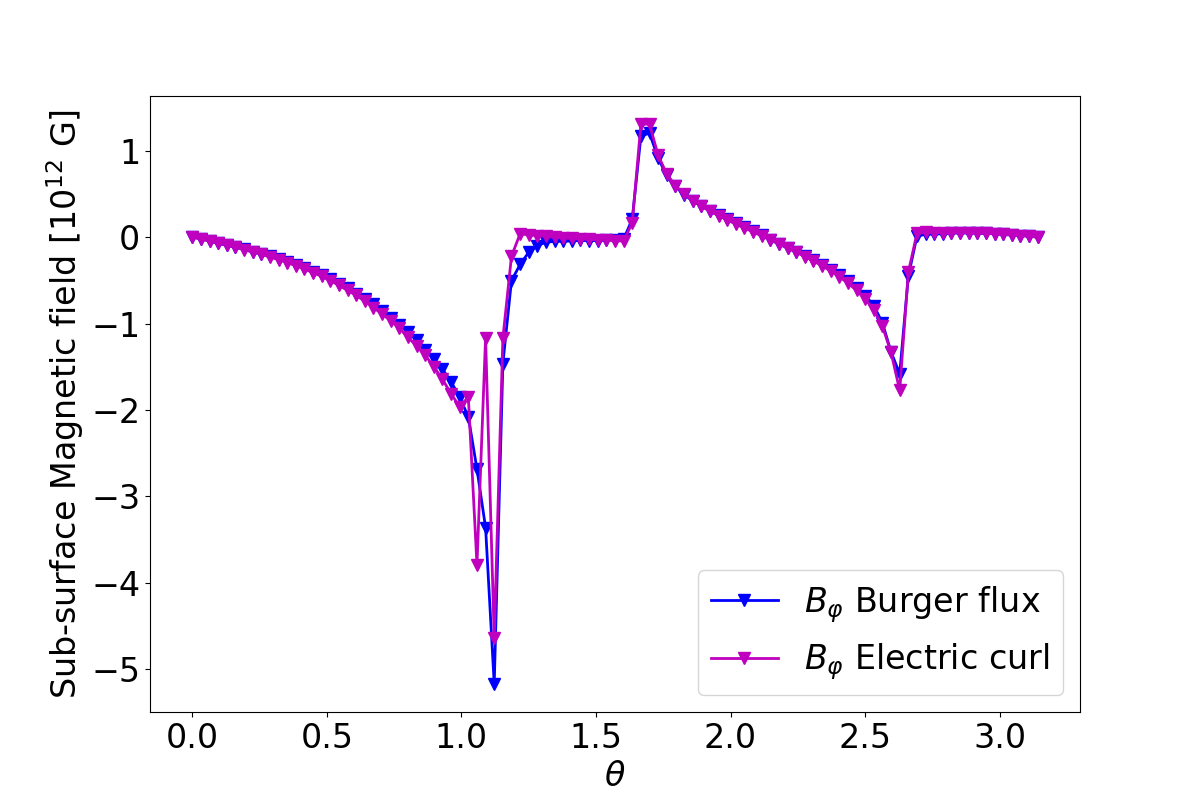}
  \caption{Example of the difference in the toroidal field advance method: profiles of $B_\theta$ (top) and $B_\varphi$ (bottom) for model {\tt crM} at $t=17$ kyr, using the Burgers-like prescription (blue) or the electric field circulation (magenta). In both cases, the poloidal magnetic field has been advanced with the upwind-like scheme, to enhance the role of the toroidal field advance method. In this specific case, we used $N_r=100$ and $N_\theta=49$, and the EULA scheme with $k_{\rm cour}=0.5$.}
  \label{fig:multipolar_burgers_comparison}
\end{figure}

The second choice is crucial to resolve the discontinuities that appear due to the Hall term in the crustal induction equation. As a typical example, we show in Fig.~\ref{fig:multipolar_burgers_comparison} the comparison of $B_\theta(\theta)$ and $B_\varphi(\theta)$ just below the surface, for the choices a. (magenta) and b. (blue), for model {\tt crM}, at an illustrative time of 17 kyr. Whenever magnetic discontinuities are created, the Burgers-like approach is able to resolve them and maintain a clean profile. On the other hand, choice a. implies a noisy and oscillating profile, which gives rise to a spurious current and electric field, ultimately affecting also the local temperature (by artificial extra Joule heating). Note that such oscillations are not due to Courant-violation instability (the numerical solution converges to the one shown for different $dt_b$): they are indeed caused by the spatial discretization scheme.

\subsection{Poloidal magnetic field advance}

The advance of the poloidal field, which in 2D is given by $\vec{B}_{\rm pol} = B_r~\hat{r} + B_\theta~\hat{\theta}$, is performed by means of the simple toroidal vector potential evolution equation:

\begin{equation}
\frac{\partial A_\varphi}{\partial t} = -c~e^\nu E_\varphi~,
\end{equation}
so that the two poloidal field components are obtained at each timestep by applying the Stokes' theorem on $A_\varphi$ with a surface $S=S_r$ or $S=S_\theta$:\footnote{Note that the line integral operator and the time advance operators numerically commute, so that if we directly evolve the $\vec{B}_{\rm pol}$ components we obtain the same results at a round-off level. However, $A_\varphi$ is a useful quantity (for instance, to draw the magnetic field lines and to apply boundary conditions), so evolving it directly avoids its reconstruction.}
\begin{equation}
\int_{S} (\vec{B}\cdot \hat{n}) ~dS = \oint_{\partial S} A_{\varphi} ~dl_\varphi ~.
\label{eq:induction_poloidal}
\end{equation}
The toroidal electric field $\vec{E}_{\rm tor}=E_\varphi~\hat{\varphi}$ is given by
\begin{equation}
  \vec{E}_{\rm tor} = \frac{\vec{j}_{\rm tor}}{\sigma_e}
  + \frac{1}{c e n_e}\vec{j}_{\rm pol}
  \times \vec{B}_{\rm pol}~,
\end{equation}
where, for each cell $(i,j)$, we consider two options to define $\vec{B}_{\rm pol}$:

\begin{itemize}
	\item[c.] A centered scheme, simply using the local values $B_r=B_r^{(i,j)}$ and $B_\theta=B_\theta^{(i,j)}$.
	\item[d.] An upwind-like scheme, assessing the poloidal electron velocity $\vec{v}_{\rm pol}\equiv-\vec{j}_{\rm pol}/(en_e)$: for instance, if $j_r^{(i,j)}>0$ (i.e., negative radial velocity), then $B_\theta=B_\theta^{(i,j+1)}$, and if $j_\theta^{(i,j)}>0$  then $B_r=B_r^{(i+1,j)}$. Note that normally upwind methods are accompanied by reconstruction methods (e.g., minmod in \cite{vigano12a}); in our case, we instead simply take the value of the field already defined and evolved at the upwind interface of the cell centered at $(i,j)$.
\end{itemize}

\begin{figure}[t!]
  \centering
  \includegraphics[width=0.45\textwidth]{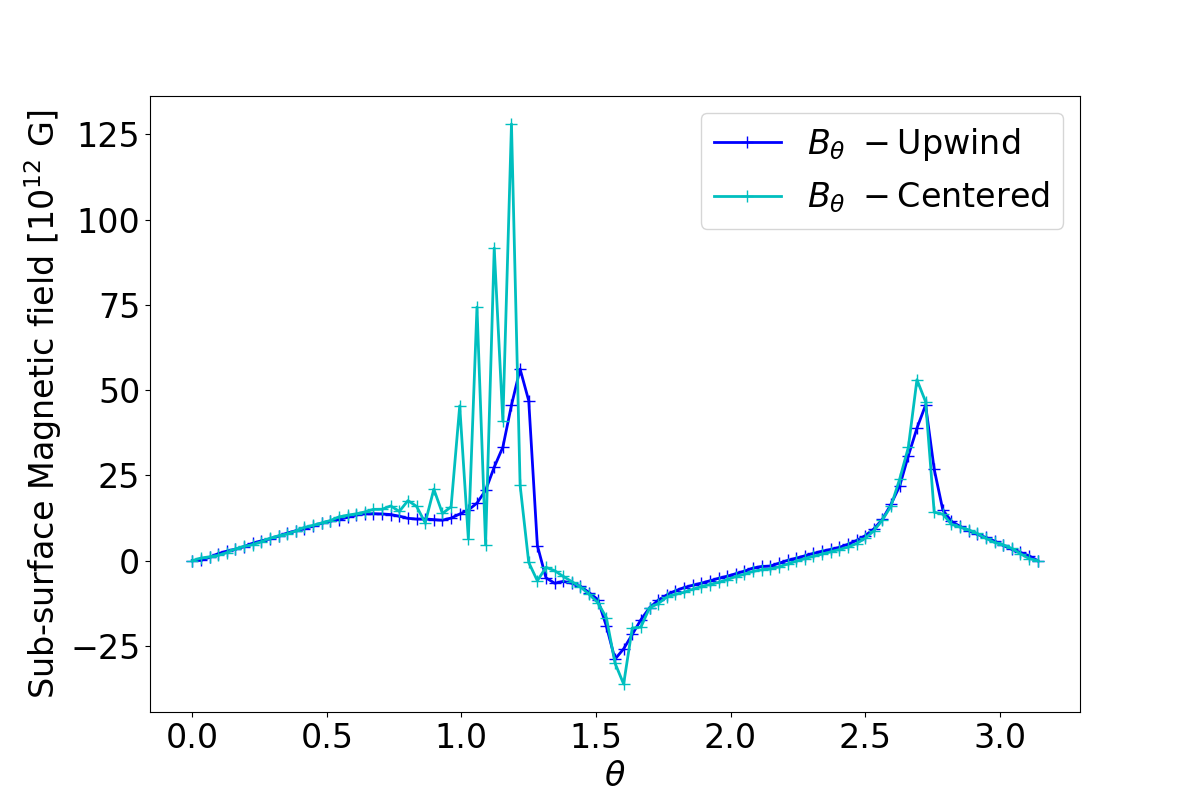}
  \includegraphics[width=0.45\textwidth]{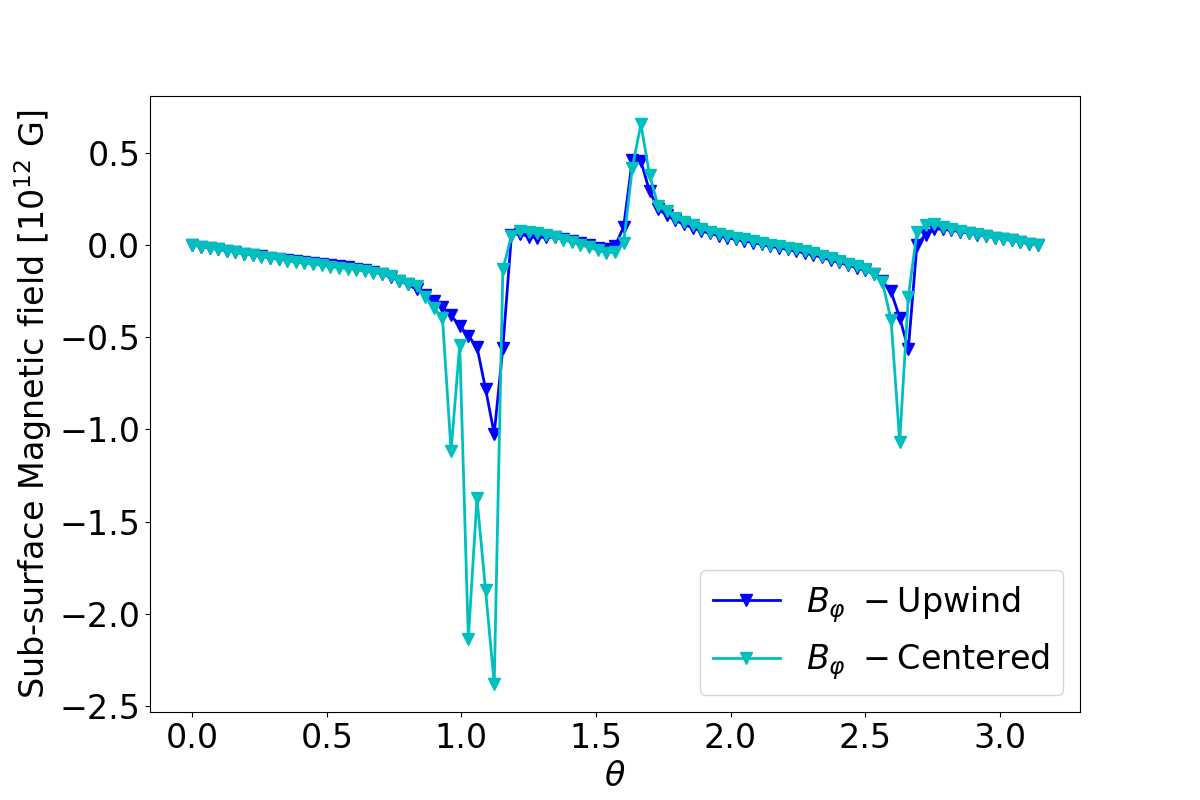}
  \caption{Example of the difference in the poloidal field advance method: profiles of $B_\theta$ (top) and $B_\varphi$ (bottom) for model {\tt crM} at $t=80$ kyr, using the upwind  prescription (blue) or the centered scheme (cyan). In both cases, the toroidal magnetic field has been advanced with the Burgers-like scheme, to enhance the role of the poloidal field advance method. Resolution and time advance method are as in Fig.~\ref{fig:multipolar_burgers_comparison}.}
  \label{fig:multipolar_etor_comparison}
\end{figure}

As before, the second choice offers a much better accuracy in the presence of discontinuities. In Fig.~\ref{fig:multipolar_etor_comparison} we compare, as an explanatory case, the tangential magnetic field meridional profiles in the crust, just below the surface, for model {\tt crM} at a late stage, $t=80$ kyr. The choice c. (cyan) is contaminated by strong oscillations, which, as above, provide artificial extra currents and unphysical additional heating (to which, as above, the solution converges numerically if the timestep is changed, thus discarding a Courant-violation origin). Instead, the choice d. (blue) offers a very clean profile, maintaining and resolving all the discontinuities.

\subsection{Hyper-resistivity}

In order to further reduce the appearance of numerical noise, we consider the application of an explicit hyper-resistive term in the $\varphi$-component of the induction equation. Its aim is ideally to dissipate the shortest spurious waves (wavelength $\gtrsim$ grid size, where numerical instabilities usually appear), without changing the global solution. Since the magnetic field is divergence-less, the following identity holds: $\nabla^2 \vec{B} = \vec{\nabla}\times(\vec{\nabla}\times\vec{B})$. We then consider two possible operators based on fourth-order derivatives.

The first one is to apply four times the Stokes operator ${\cal S}$ to $B_\varphi$, so that:
	\begin{equation}
	\partial_t \vec{B}_{\rm tor} \rightarrow \partial_t \vec{B}_{\rm tor} -\eta_{\rm curl4}\Delta^2{\cal S}^4 \vec{B}_{\rm tor}~.
	\end{equation}
	Note that ${\cal S}$ applied on the value $B_\varphi^{(i,j)}$ includes only the first neighbors in each direction, $(i\pm 1,j\pm 1)$. Therefore, this operator is able to smooth out oscillations down to a minimum scale of twice the magnetic grid size.

The second possibility is to apply twice a finite-difference vector Laplacian operator to the toroidal field
	\begin{equation}
	\partial_t \vec{B}_{\rm tor} \rightarrow \partial_t \vec{B}_{\rm tor} -\eta_{\rm lapl2}\Delta^2\nabla^2(\nabla^2 \vec{B}_{\rm tor})~,
	\end{equation}
where $\nabla^2 \vec{B}_{\rm tor}$ includes first and second-order derivatives, which are evaluated by standard second-order accurate centered formulae. Therefore, it couples 5 points, odd and even, in each direction and is able to damp oscillations of the grid size.

In both cases, the pre-coefficient includes the grid size squared $\Delta^2 = [1/dr^2 + 1/(r~d\theta)^2]$, and a dimension-less free parameter $\eta_{\rm curl4}$ or $\eta_{\rm lapl2}$. In order to avoid changing the global solution (and not to cause further restrictions to the timestep), typically we found $\eta_{\rm curl4} \lesssim 0.1$ and $\eta_{\rm lapl2} \lesssim 0.001$, for which the additional numerical dissipation of energy is not more than a few percent in $\sim 10^5$ yr (after several millions of time steps).

The use of hyper-resistivity helps stabilizing the code especially for fine resolutions or high initial magnetic fields at late times ($\gtrsim 10^5$ yr), when the star is cold and the Hall term dominates. The explicit hyper-resistivity can be applied in combination with any of the other space and time discretization methods detailed above. In the simulations shown in this paper, we do not apply it.

\section{Numerical and computational analysis}
\label{sec:computational}

We now analyze the methods outlined above and our implementation from various points of view: numerical convergence, energy conservation, stability, and a theoretical computational complexity analysis including a performance study of the most important blocks of calculation.

\subsection{Convergence}

\begin{figure}[!t]
  \centering
  \includegraphics[width=0.45\textwidth]{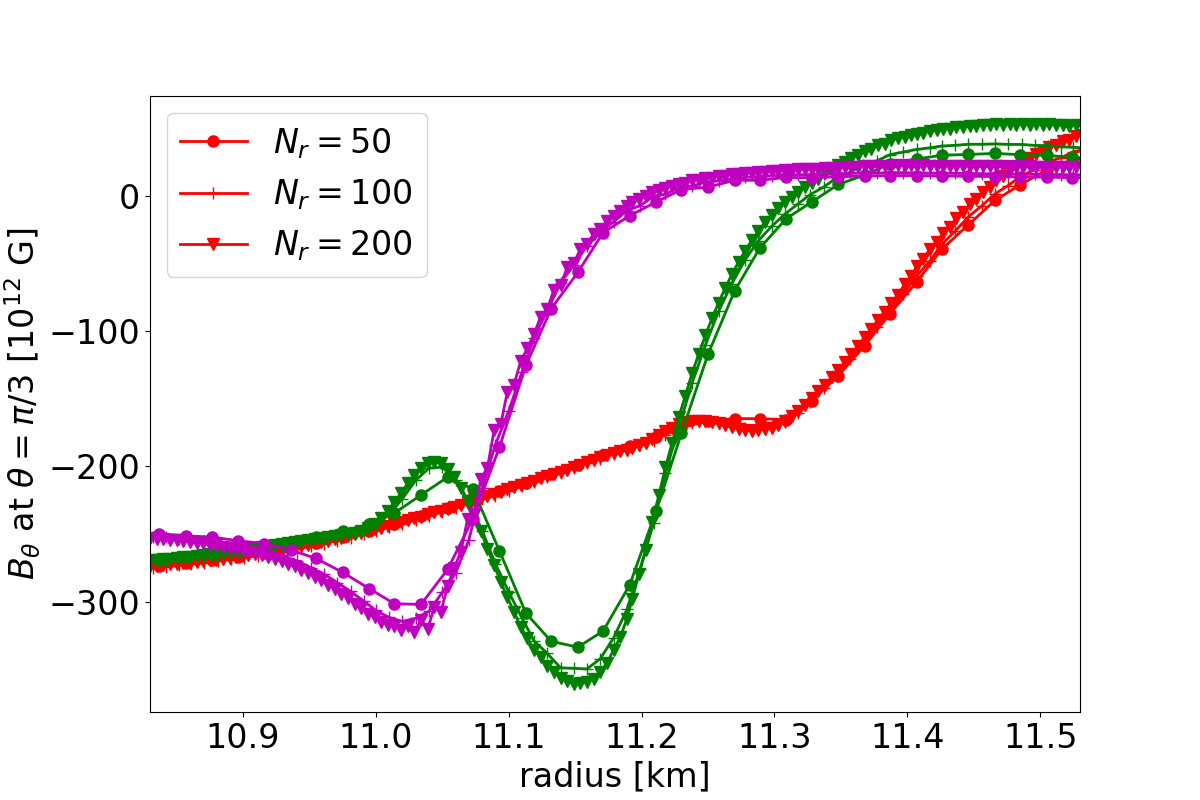}\\
  \includegraphics[width=0.45\textwidth]{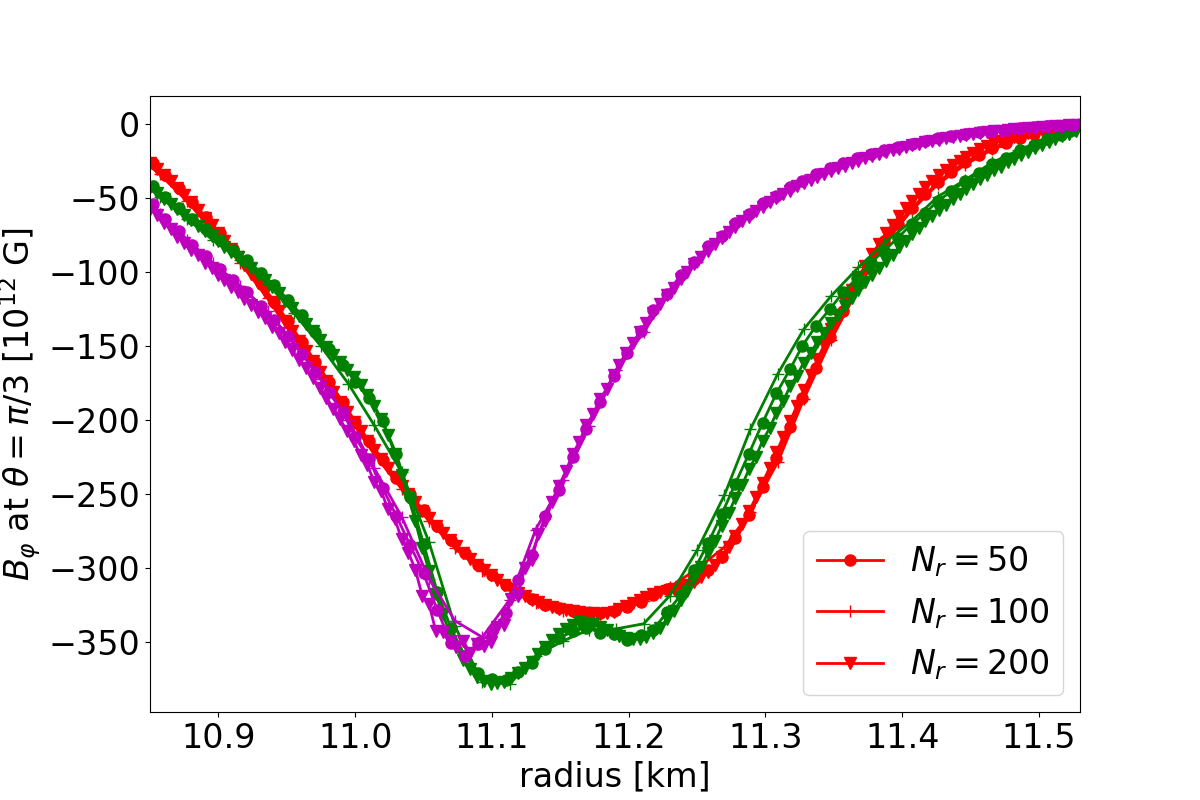}
  \caption{Convergence with numerical radial resolution $N_r=50,100,200$, with $N_\theta=49$ (number of thermal cells): radial profile of $B_\theta$ (top) and $B_\varphi$ (bottom) for model {\tt crM}, at 1, 10 and 50 kyr (red, green and magenta, respectively). Here we show the EULA method, with $k_{\rm cour}=0.5$, but other methods behave similarly.}
  \label{fig:convergence}
\end{figure}

\begin{figure}[!t]
  \centering
  \includegraphics[width=0.45\textwidth]{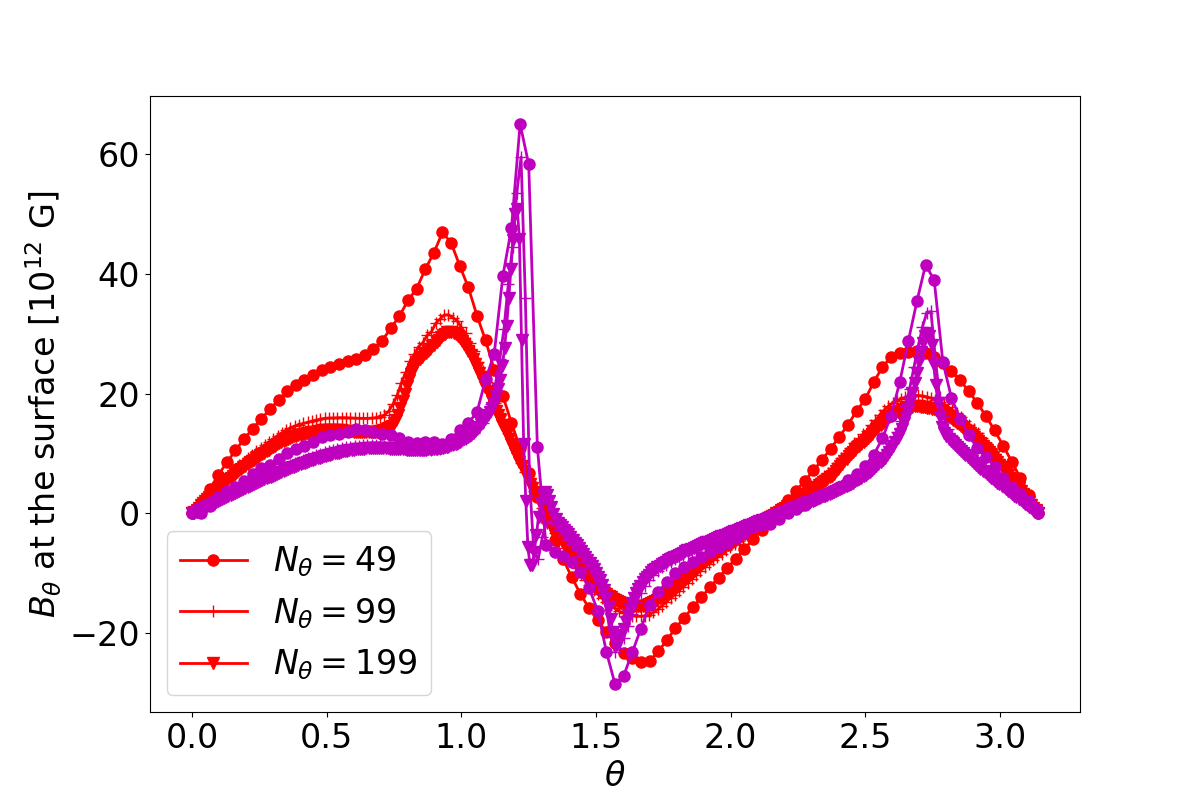}
  \caption{Convergence with numerical angular resolution $N_\theta=49,100,200$ with $N_r=100$ (number of thermal cells): meridional profile of $B_\theta$ at the surface for model {\tt crM}, at 1 and 50 kyr (red and magenta respectively), with the same numerical methods as Fig.~\ref{fig:convergence}.}
  \label{fig:convergence_angular}
\end{figure}

In Fig.~\ref{fig:convergence} we show as an example the convergence of the numerical solution as we increase the radial resolution, in this case for model {\tt crM} evolved with the EULA method. The radial profile of any component of $\vec{B}$, $\vec{j}$ and $\vec{E}$ (we show here the representative case $B_\theta$ and $B_\varphi$ for model {\tt crM} at different times) for $N_r=100$ and 200 are very close to each other and resolve better the regions with the largest gradients, compared with the case $N_r=50$. The meridional resolution behaves similarly, as shown in Fig.~\ref{fig:convergence_angular}, which shows the meridional profile of $B_\theta$ at the surface, close to which the largest differences are seen.\footnote{Note that the application of the boundary condition for the magnetic field requires in general the integral over the surface of quantities involving $B_r(\theta)$ or $A_\varphi(\theta)$ (see \cite{pons19}), the accuracy of which depends on the angular resolution. Therefore, the meridional resolution affects not only the capability of resolving the eMHD dynamics, but also the reconstruction of the $B_\theta$ corresponding to the potential solution at the boundary.}

\begin{figure}[!t]
	\centering
	\includegraphics[width=0.45\textwidth]{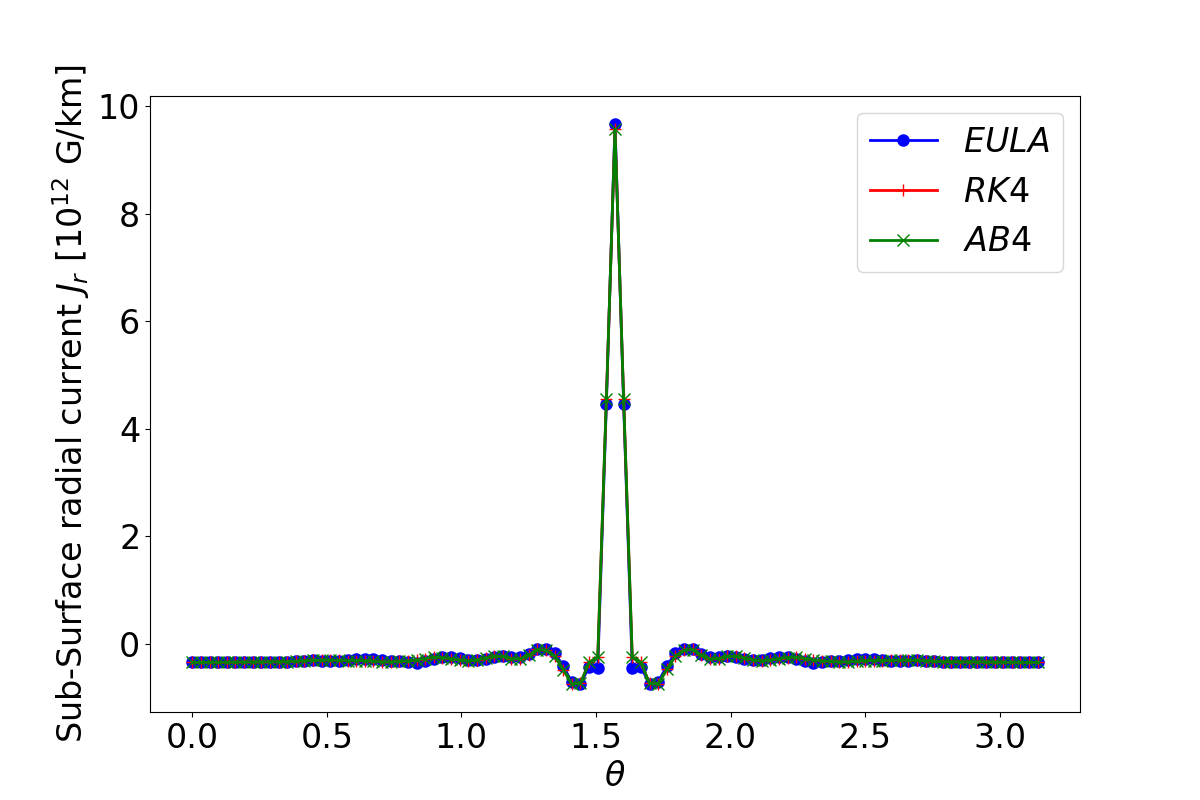}
	\caption{Comparison of different methods in their ability to resolve the current sheet (we show $j_r$ two points below the surface) for model {\tt CrP}, at $t=3$ kyr, for EULA (blue), RK4 (red) and AB4 (green): they all basically overlap. All of them use the Burgers-like and upwind-like formulations.}
	\label{fig:current_sheet}
\end{figure}

In general, RK4, AB4 and the EUL all converge to the same numerical solution for small-enough timesteps. Instead, the EULA method evolves with slight differences, which are more evident close to the surface and tend to decrease with spatial resolution. This is due to the fact that the EULA scheme corresponds to the introduction in the equations of a hyper-resistivity term in the poloidal field evolution, so that the discretized equations result to be slightly different. In Fig.~\ref{fig:current_sheet} we show how the numerical methods perform in resolving the naturally arising current sheets, like the ones that develop at the equator for model {\tt CrP}. The plot shows the meridional profile of $j_r$ two points below the surface. Note that spectral methods would not be able to resolve such sharp peaks in currents (i.e., large discontinuities in the magnetic field components).

In general, the largest differences with resolution and between the different finite-volume methods tend to appear mostly in the outermost layers during the initial transient phase (lasting some centuries) and disappear soon after.

Besides the magnetic field details shown here, the global quantities (dipolar component at the surface, magnetic energy, integrated Joule heat, luminosity) converge to the same result. A resolution $N_\theta=49$, $N_r=100$ is already able to capture the most important features.

Note that a more quantitative assessment of the convergence order and the accuracy are basically unfeasible. As a matter of fact, the dependence of the result non-linearity of the problem and the interplay between magnetic field, microphysical coefficients and temperature is such that: (i) no analytical solutions are available, and (ii) the dependence of a given local or global quantity with resolution is non-trivial and time-dependent. Anyway, even for a simplified problem where such analysis could be available (for instance, constant temperature, microphysical coefficients and a simplified magnetic topology), we would not expect the convergence order to be higher than 1, due to the spatial scheme used. On the other hand, higher-order spatial schemes would likely cause additional instabilities (see \cite{vigano19}).

\subsection{Energy conservation}

The fulfillment of energy balance \cite{pons09} can be evaluated by looking at how well the total energy is conserved in time:

\begin{equation}
E_{\rm tot}(t) = E_{\rm mag}(t) + \int_0^t Q_j~dt' + \int_0^t P_{\rm out}~dt'~,
\end{equation}

where $E_{\rm mag}=\int_{V_\star} e^\nu (B^2/8\pi)~dV_\star$ is the magnetic energy stored in the star, $Q_j = \int_{V_\star} (e^{2\nu} j^2/\sigma_e) dV_\star$ is the volume-integrated (positive definite) Joule dissipation rate, and $P_{\rm out}=(1/4\pi)\int_{S_\star} [e^{2\nu}(\vec{E}\times\vec{B})\cdot\hat{r}]~dS_\star$ is the outgoing Poynting flux integrated over the outermost cell interfaces (surface $S_\star$).

The decrease in magnetic energy is caused by the Joule dissipation. The Poynting flux across the surface is usually negative, as a direct result of the internal dissipation: as the poloidal field decreases inside, the magnetospheric field also shrinks in response, thus causing a gradual loss of magnetic energy stored in the magnetosphere.

\begin{figure}[!t]
	\centering
	\includegraphics[width=0.45\textwidth]{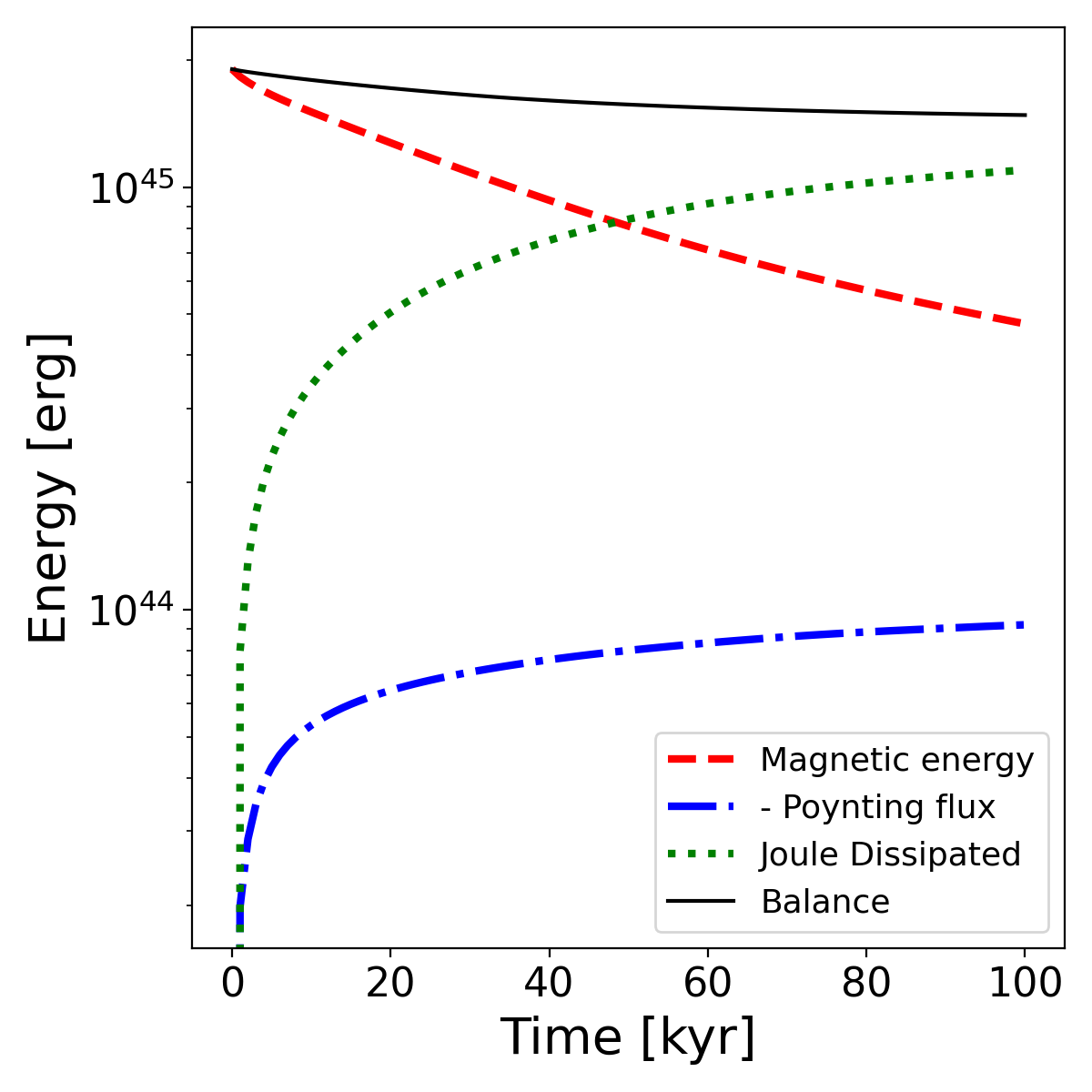}
	\caption{Conservation of energy for model {\tt crM} with $N_r=100$ and $N_\theta=49$, EULA method (the same plot holds for the other finite-volume methods), upwind and Burgers-like schemes: total energy (Balance, solid line), star's magnetic energy (red dashes), time-integrated incoming Poynting flux (blue dot-dashed line), time-integrated Joule dissipated heat (green dots). The numerical total energy loss after 100 kyr in this case is of about $22\%$.}
	\label{fig:conservation}
\end{figure}

In Fig.~\ref{fig:conservation}, we show the numerical results for model {\tt crM}. The numerical loss of the total integrated energy, $1-E_{\rm tot}(t)/E_{\rm mag}(t=0)$, is in the range of $18-27\%$ at 100 kyr, for resolutions $N_r,(N_\theta+1)\in [50,200]$, where finer resolution allows a better conservation. Besides the finite resolution, part of the numerical energy loss comes from the approximations made at the crust-core interface (see \S~\ref{sec:cc_interface}). When hyper-resistivity is included, the numerical dissipation adds an additional energy loss of a few percent at most.

\subsection{Methods: stability and optimum timestep}

\begin{table}[!t]
	  \centering
	\begin{tabular}{c c c c }
		\hline
		\hline
		method & $T$ [K] & points & $k_{\rm opt}$ \\
		\hline
		EULA & evo. & $99 \times 75$ & 0.8 \\
		RK4 & evo. & $99 \times 75$ & 0.3 \\
		AB4 & evo. & $99 \times 75$ & 0.2 \\
		\hline
		EULA & $10^9$ & $99 \times 75$ & 0.8 \\
		RK4 & $10^9$ & $99 \times 75$ & 2.3 \\
		RK4 & $10^9$ & $199 \times 150$ & 1.7 \\
		AB4 & $10^9$ & $99 \times 75$ &	0.2 \\
		EUL & $10^9$ & $99 \times 75$ & 0.1 \\
		\hline
		EULA & $10^8$ & $99 \times 38$ & 0.9 \\
		EULA & $10^8$ & $99 \times 75$ & 0.8 \\
		EULA & $10^8$ & $99 \times 150$ & 0.8 \\
		RK4 & $10^8$ & $99 \times 75$ & 0.1 \\
		AB4	& $10^8$ & $99 \times 75$ &	0.2 \\
		EUL & $10^8$ & $99 \times 75$ & $<0.02$ \\
		\hline
		\hline
	\end{tabular}
	\caption{Optimal value $k_{\rm opt}$ that guarantees stability up to 10 kyr, for different methods for model {\tt CrP}, with temperature fixed or evolving as indicated. The resolution indicates the angular times radial number of points in the crust where magnetic field is evolved.}
	\label{tab:kc}
\end{table}

Generally speaking, numerical instabilities in eMHD magneto-thermal simulations are prone to appear especially during two stages: (i) during the first centuries due to the fast transient waves associated with the out-of-equilibrium initial conditions; (ii) when the magnetization parameter $f_hB/\eta$ exceeds $\gtrsim 100$, either because of very strong fields $B\gtrsim 10^{15}$ G, or because the  conductivity becomes relatively high when the star cools below $T\lesssim 10^8$ K, which happens around the switch from the neutrino-dominated era to the photon-dominated era (${\cal O}(10^5)$ yr).
The instabilities during the first stage tend to appear in the outermost layers of the crust, where $f_h$ is larger. It represents a caveat against the quantitative meaningfulness of results for very young stars, but it is usually transient (the resistivity is high and tends to damp short and fast waves) and does not affect the results at observationally meaningful ages $\gtrsim$ kyr. On the other hand, the late-stage instabilities can be reflected in artificial bumps in the calculated temperature map and luminosity, caused by the Joule heating associated with the perturbations combined with the fact that the neutron star's temperature and heat capacity are greatly reduced. This is one of the factors limiting the validity of the simulations at times $\gtrsim 10^5-10^6$ yr.

Our methods aim at extending as much as possible the range of feasibility of the simulation, in terms of stability and computational time. We define $k_{\rm opt}$ as the maximum value of $k_{\rm cour}$, defined in eq.~(\ref{eq:dtb_kc}), that allows numerical stability (defined as absence of noise in the magnetic field profiles), for a given configuration and method. We show the results in Table~\ref{tab:kc}, where we have compared the solutions of model {\tt CrP} up to $10^4$ yr. We have considered different resolutions, and either a fixed $T=10^8$ or $10^9$ K, or the full magneto-thermal evolution. The largest values of $k_{\rm opt}$ are found for EULA and RK4. However, RK4 suffers from a few problems: (i) $k_{\rm opt}$ decreases notably for lower temperatures (i.e., higher magnetization) and (ii) for higher resolution; (iii) it is in general slower than EULA because, even in the case where $k_{\rm opt}$ is larger for RK4, each timestep contains four sub-steps. The other methods (AB4 and EUL) are much slower, having small values of $k_{\rm opt}$ and much longer CPU time. Note that in all cases the introduction of the hyper-resistivity can at best increase only slightly the values of $k_{\rm opt}$.

Moreover, for all methods $k_{\rm opt}$ becomes much smaller when later ages (i.e., temperatures below $10^8$ K) are considered. As a matter of fact, in both cases (early and late instabilities), the trigger is numerical and a typical signature is the high-frequency noise in the profile of the magnetic field components, breaking any topological symmetry analytically expected (for instance, in the evolution of a pure dipole). The non-linearity of the equations makes them grow, unless the resistive terms cure it. This is why the late-time instabilities for high magnetization parameters are intrinsically harder to be cured, and lowering the timestep may not be enough.

We conclude that the EULA method is the most efficient one, being the fastest one and the only one showing a value of $k_{\rm opt}\sim 0.8$ almost independent of the resolution and of the model; a choice $k_{\rm cour} \lesssim k_{\rm opt}$ is advised for simulations up to $\lesssim 10^5$ yr. If later ages $t \lesssim 10^6$ yr are considered, $k_{\rm cour}\sim 0.1$ and hyper-resistivity should be used.

\begin{table}[!t]
	\centering
	\resizebox{\linewidth}{!}{
		\begin{tabular}{|c|c|c c|c c|}
			\hline
			& & \multicolumn{4}{|c|}{Runtime~$[\%]$}\\
			\hline
			Block & Complexity & \multicolumn{2}{|c|} {\tt Core} & \multicolumn{2}{|c|} {\tt CrP}\\
			& & (1 kyr) & (100 kyr) & (1 kyr) & (100 kyr ) \\
			\hline
			Conductivities & $\mathcal{O}(N_r N_\theta)$ & $16.2$ & $14.7$ & $50.4$ & $6.7$\\
			Thermal Evolution & $\mathcal{O}( N_r N_\theta^3)$ & $46.5$ & $10.2$ & $32.6$ & $3.4$\\
			Heat Capacity & $\mathcal{O}(N_r N_\theta)$ & $2.2$ & $1.4$ & $3.1$ & $0.7$\\
			Neutrino Emissivity & $\mathcal{O}(N_r N_\theta )$ & $3.4$ & $0.8$ & $3.7$ & $0.8$\\
			Magnetic Evolution & $\mathcal{O}(N_\theta N_r) $ & $28.5$ & $49.1$ & $8.6$ & $74.8$\\
			\hline
		\end{tabular}}
		\caption{Summary of the complexity analysis for the most computationally costly subroutines showing the asymptotically worst-case complexity for each one of the subroutines and the fraction of runtime they take to run models {\tt CrP} and {\tt Core}, up to 1 or 100 kyr, with $N_r=100$ and $N_\theta=49$, EULA scheme, $k_{\rm cour}=0.5$, with upwind and Burgers-like treatment}. 
		\label{tab:complexity}
	\end{table}

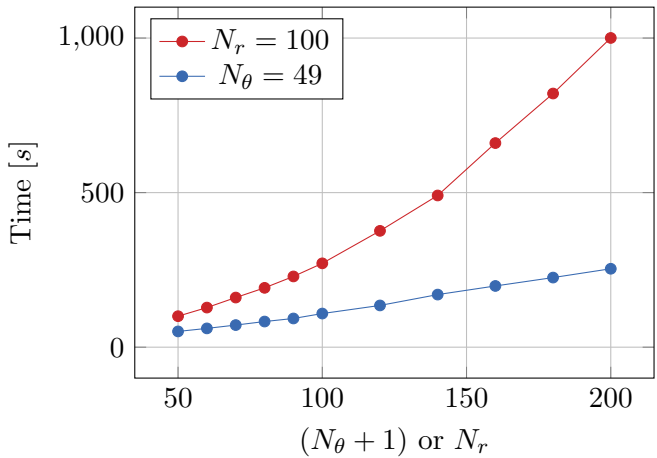
\begin{figure}[!t]
	\begin{tikzpicture}
	\begin{axis}[
	xlabel={$ (N_\theta+1) $ or $N_r$},
	ylabel={Time~$[s]$},
	height=6.5cm,
	width=0.95\linewidth,
	grid=major,
	ymin=0,
	ymax=1000,
	xmin=50,
	xmax=200,
	enlargelimits=true,
	yticklabel style={
		/pgf/number format/fixed,
	},
	legend pos=north west,
	scaled ticks=false]
	\addplot[nicered, mark=*,mark options={fill=nicered}] table [x=kmax,y=mt2d]{data/mt2d_kmax.txt};
	\addlegendentry{$N_r=100$}
	\addplot[niceblue, mark=*,mark options={fill=niceblue}] table [x=lmax,y=mt2d]{data/mt2d_lmax.txt};
	\addlegendentry{$ N_\theta =49$}
	\end{axis}
	\end{tikzpicture}
	\caption{Runtime of the whole simulation ({\tt CrP} model up to 1 kyr, with the same setup as in Table~\ref{tab:complexity}) as a function of $N_\theta$ for a fixed $N_r=100$ (red line) or a function of $N_r$ for a fixed $ N_\theta=49$ (blue line). We observe a linear scaling for $N_r$ and a worse, non-linear one for $N_\theta$, due to the contribution of the matrix inversion, here performed with our Thomas algorithm implementation.}
	\label{fig:performance}
\end{figure}

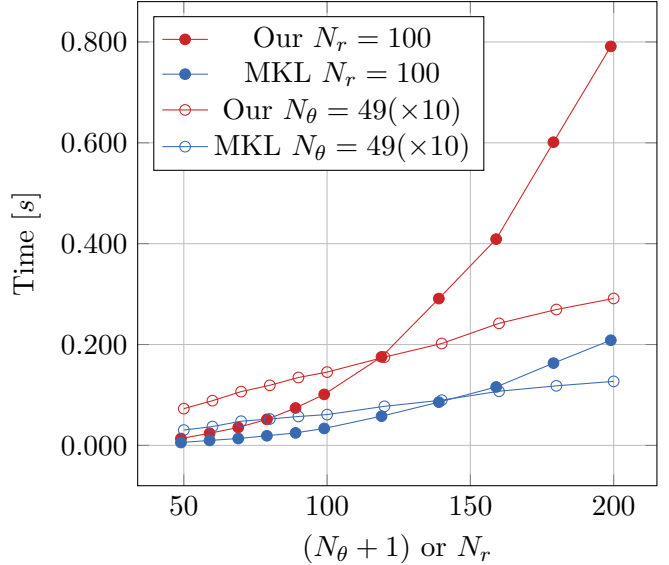
\begin{figure}[!t]
	\centering
	\begin{tikzpicture}
	\begin{axis}[
	xlabel={$(N_\theta + 1)$ or $N_r$},
	ylabel={Time~$[s]$},
	height=8cm,
	width=0.95\linewidth,
	grid=major,
	enlargelimits=true,
	ymin=0,
	ymax=0.8,
	yticklabel style={
		/pgf/number format/fixed,
		/pgf/number format/zerofill,
		/pgf/number format/precision=3
	},
	legend pos=north west]
	\addplot[nicered, mark=*,mark options={fill=nicered}] table [x=kmax,y=solvtb]{data/mkl_kmax.txt};
	\addlegendentry{Our $N_r = 100$}
	\addplot[niceblue, mark=*,mark options={fill=niceblue}] table [x=kmax,y=mkl]{data/mkl_kmax.txt};
	\addlegendentry{MKL $N_r = 100$}

	\addplot[
	nicered,
	mark=o,
	mark options={fill=nicered}
	] table [x=lmax,y expr=\thisrow{solvtb}*10]{data/mkl_lmax.txt};
	\addlegendentry{Our $N_\theta = 49 (\times 10)$}
	\addplot[
	niceblue,
	mark=o,
	mark options={fill=niceblue},
	] table [x=lmax,y expr=\thisrow{mkl}*10]{data/mkl_lmax.txt};
	\addlegendentry{MKL $N_\theta = 49 (\times 10)$}
	\end{axis}
	\end{tikzpicture}
    \caption{Runtime for the thermal evolution matrix inversion as a function of number of blocks (angular points $N_\theta$, filled points) or as a function of the diagonal inner dimension (radial points $N_r$, empty points, multiplied by 10 for better visualization). We compare our Thomas' algorithm implementation (red) and Intel's \acs{MKL} one (blue) to run one matrix solving call.}
    \label{fig:scalability_mkl}
\end{figure}

\subsection{Computational analysis}

As base models to assess the computational cost, we run models {\tt CrP} and {\tt Core}, both for 1 kyr and 100 kyr, with $N_r=100$ and $N_\theta = 49$, using the EULA scheme with $k_{\rm cour}=0.5$, the Burgers-like and upwind-like discretization schemes described above. We analyze the asymptotic computational complexity for each one of the parts that compose the main simulation loop. In Table~\ref{tab:complexity} we gather the big O notation and the fraction of computational cost of the most relevant parts (accounting together for $\sim 95\%$ of the total runtime). Most of the time is spent in the magnetic evolution in these cases: (i) in both 100 kyr runs, since on average the magnetic timestep is much smaller than the cooling timestep; (ii) at all times for the {\tt Core} model, where the elliptic equation~(\ref{eq:elliptic}) related to ambipolar diffusion is solved by the costly matrix inversion (using the same algorithm of the thermal evolution). However, the microphysics and the matrix inversion for the cooling scheme can represent the majority of the cost at the beginning of the crust-confined simulation (see {\tt CrP} up to 1 kyr), or in general for weak magnetic fields (not shown here). This is due to the fact that in those cases the magnetic and cooling timesteps are comparable, and the single computations of microphysics and thermal matrix inversion are much more costly than the magnetic evolution.

We have then analyzed the raw performance of the code. As shown in the complexity analysis, the two parameters that have the greatest impact on runtime are the grid dimensions $ N_\theta $ and $N_r$. Therefore, we have conducted a performance study of the code runtime\footnote{All experiments were run in the following test machine: Ubuntu Linux 18.04, Intel i7-4790K (4.00 GHz), 16 GiB DDR4 RAM, Samsung 840EVO SSD drive for output storage. Code was compiled with {\tt{CMake 3.0}} and {\tt{gfortran 7.5.0}} with {\tt{O3}} optimization flag enabled. Note that the implementation is single-threaded.} in terms of those two parameters within reasonable ranges: $ (N_\theta+1) , N_r \in [50, 200]$. We show the results from {\tt CrP} model up to 1 kyr in Figure \ref{fig:performance}. As we can observe, the runtime ranges between $50-800$ seconds depending on the resolution; furthermore, as expected, the computational cost has a steeper dependence on $N_\theta$ than on $N_r$.

The reason for the worse scalability of $ N_\theta $ is related to the matrix inversion algorithm in the thermal evolution. Since the matrix inversion has an important weight in the computational cost of a simulation, we have tested two ways, numerically equivalent at round-off level: (a) the manual implementation of the standard Thomas algorithm relying on the LU decomposition\footnote{The Thomas algorithm is optimized if the dimension of each block is less than the number of blocks, i.e., if the dimension with less points (usually $\theta$, i.e., $N_\theta$) is swept in the inner diagonals, and the blocks sweep the more numerous dimension ($N_r$).}; (b) the Intel \ac{MKL} \cite{wang14} implementation, which features highly optimized, threaded, and vectorized math functions that maximize performance on each processor family (which, in this case, treats the block tridiagonal matrix as a band matrix and solves it by calling {\tt{LAPACK}} subroutines for factoring and solving band matrices following a custom version of the Thomas' algorithm too). The latter shows a significant reduction in the computational cost of the whole matrix solving calls (as we can observe in Fig. \ref{fig:scalability_mkl}, the \acs{MKL} implementation scales better with speedup factors of $\sim \times 2-5$ depending on the diagonal's inner dimension and the number of blocks). On the other hand, such libraries may not work out of the box for every system and performance may differ if Intel processors are not used (and even between different families of Intel hardware), so that the well-known Thomas algorithm can be coded from scratch.

\section{Case studies}
\label{sec:casestudies}

\subsection{Crustal-confined multipolar initial field}
\label{subsec:multipolar}

\begin{figure}[!h]
	\centering
	\includegraphics[width=0.9\linewidth, clip=true]{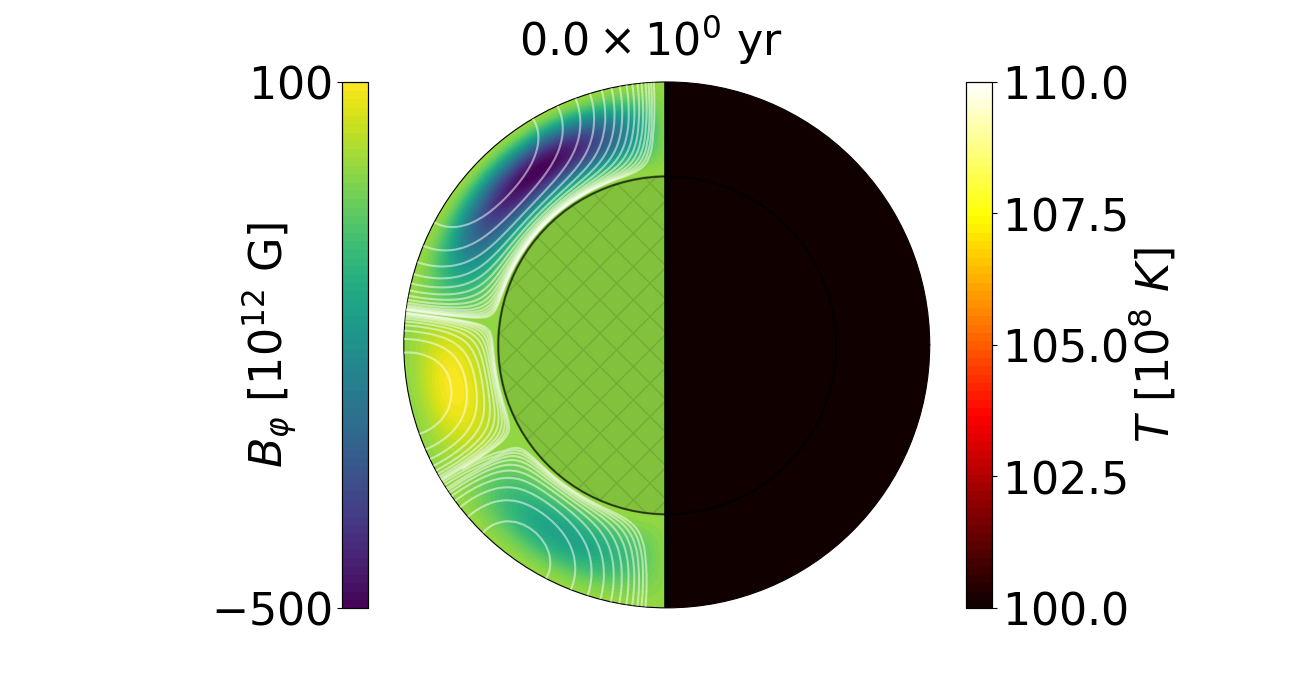}\\
	\includegraphics[width=0.9\linewidth, clip=true]{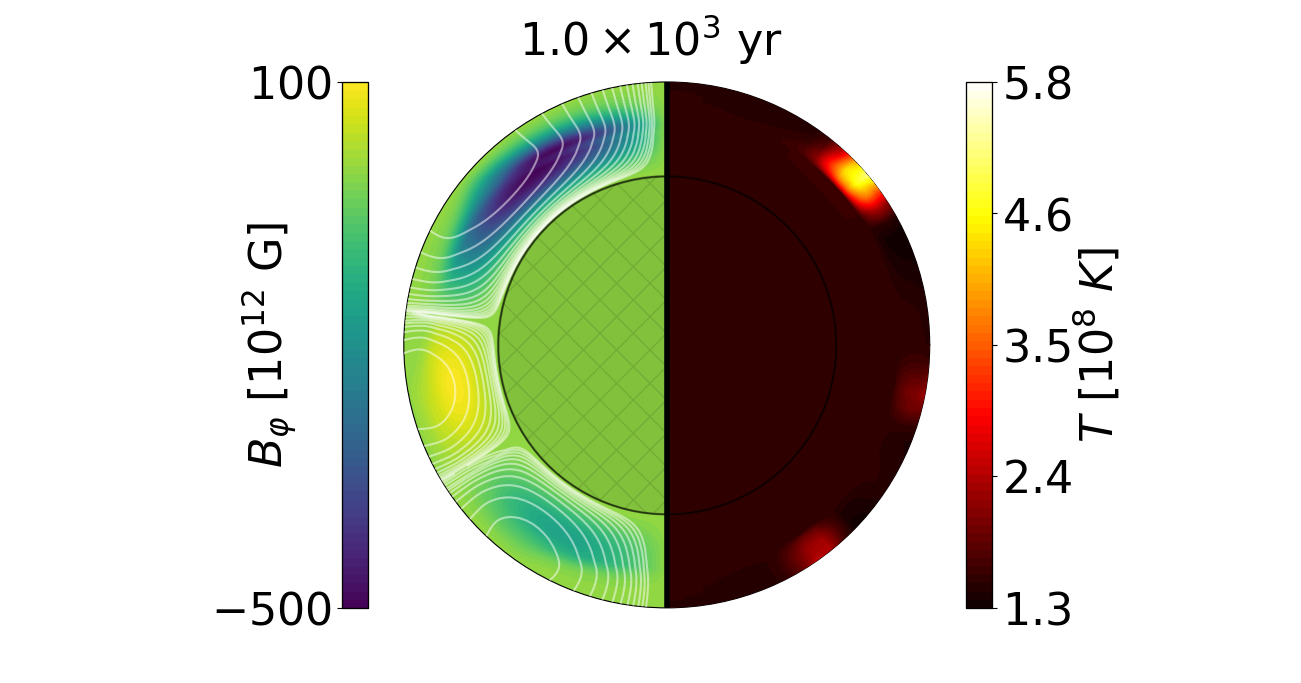}\\
	\includegraphics[width=0.9\linewidth, clip=true]{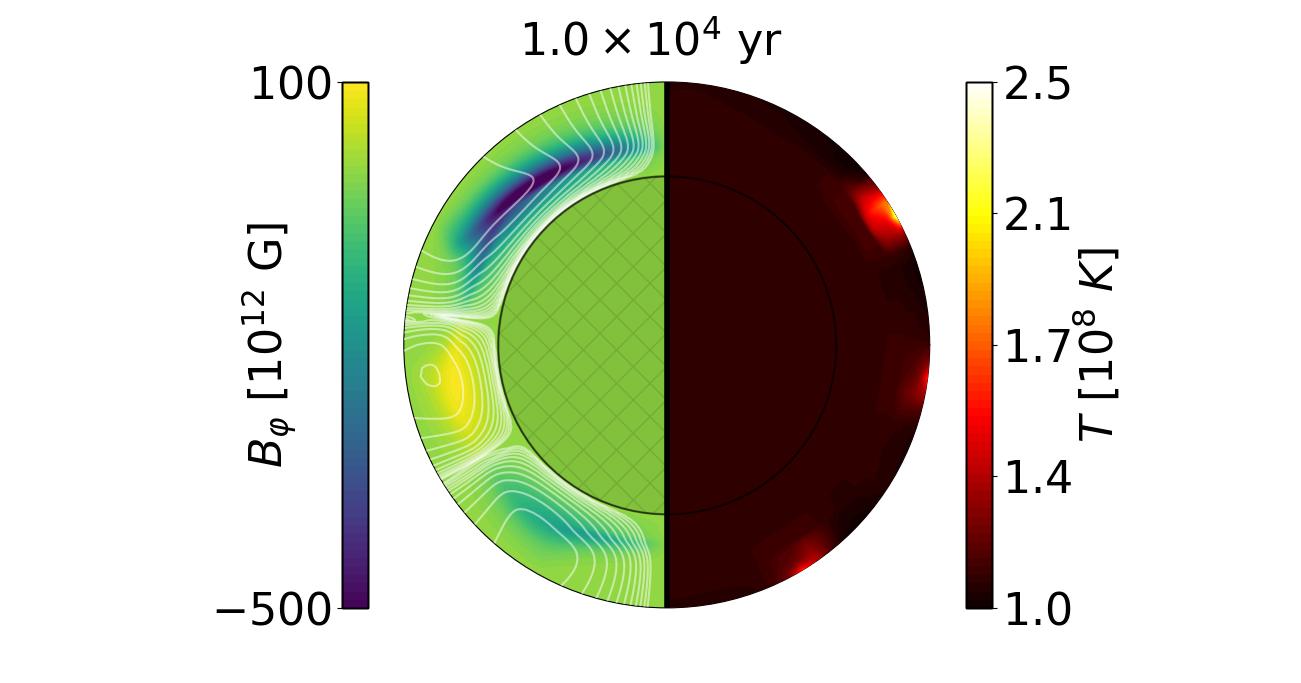}\\
	\includegraphics[width=0.9\linewidth, clip=true]{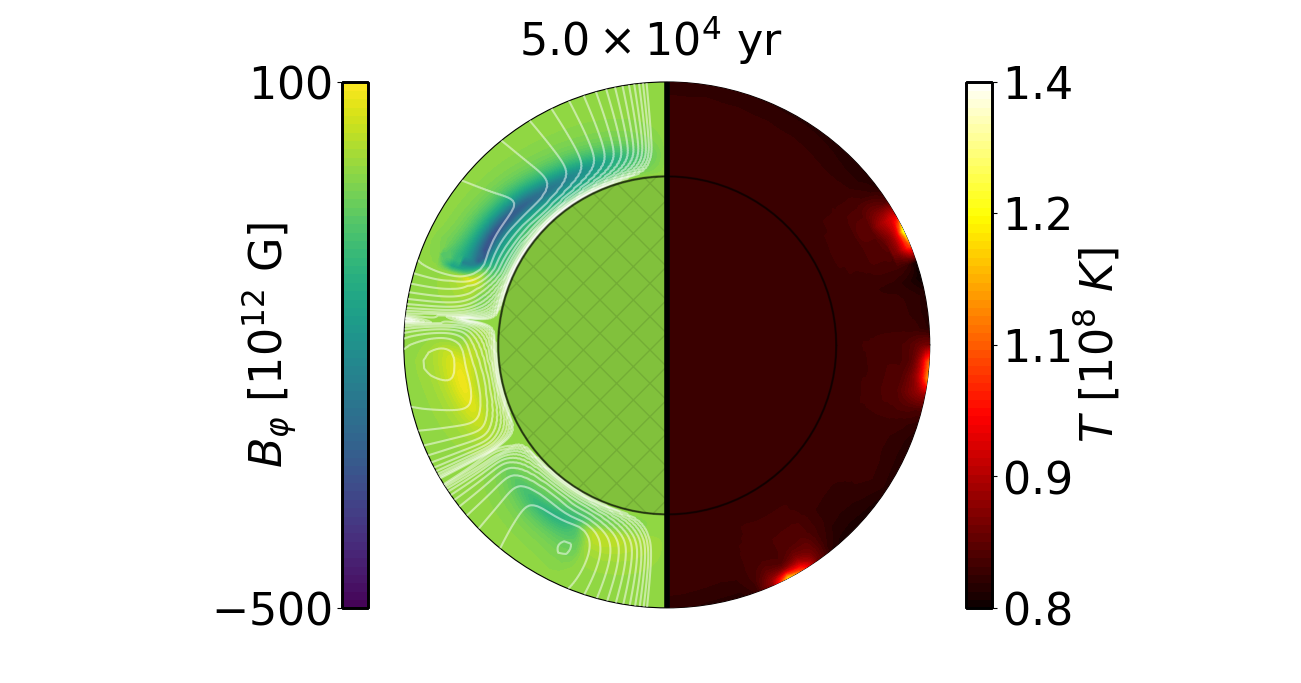}
	\caption{Evolution of magnetic field and temperature for the crust-confined multipolar model {\tt crM}, showing the meridional projection of the magnetic field lines (white lines) and the toroidal field (colors) on the left, and the internal temperature distribution (right), at  $t=0,1,10,50$ kyr. The crust has been enlarged by a factor 8 for visualization purposes. We use the optimal methods (EULA with $k_{\rm cour}=0.5$, upwind and Burgers'-like schemes), and a resolution of $N_r=100$, $N_\theta=49$. The crust has been enlarged a factor 8 for the sake of clarity.}
	\label{fig:multipolar_evolution}
\end{figure}

\begin{figure}[h!]
	\centering
	\includegraphics[width=\linewidth]{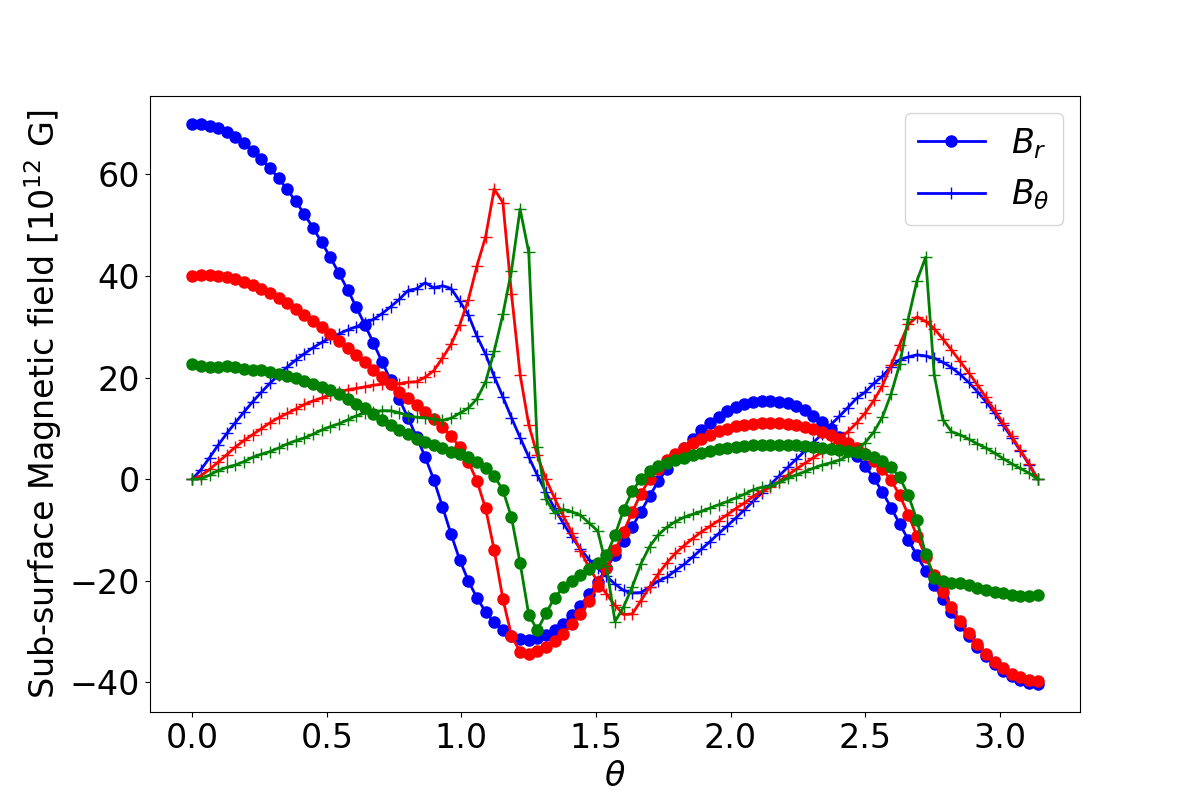}
	\includegraphics[width=\linewidth]{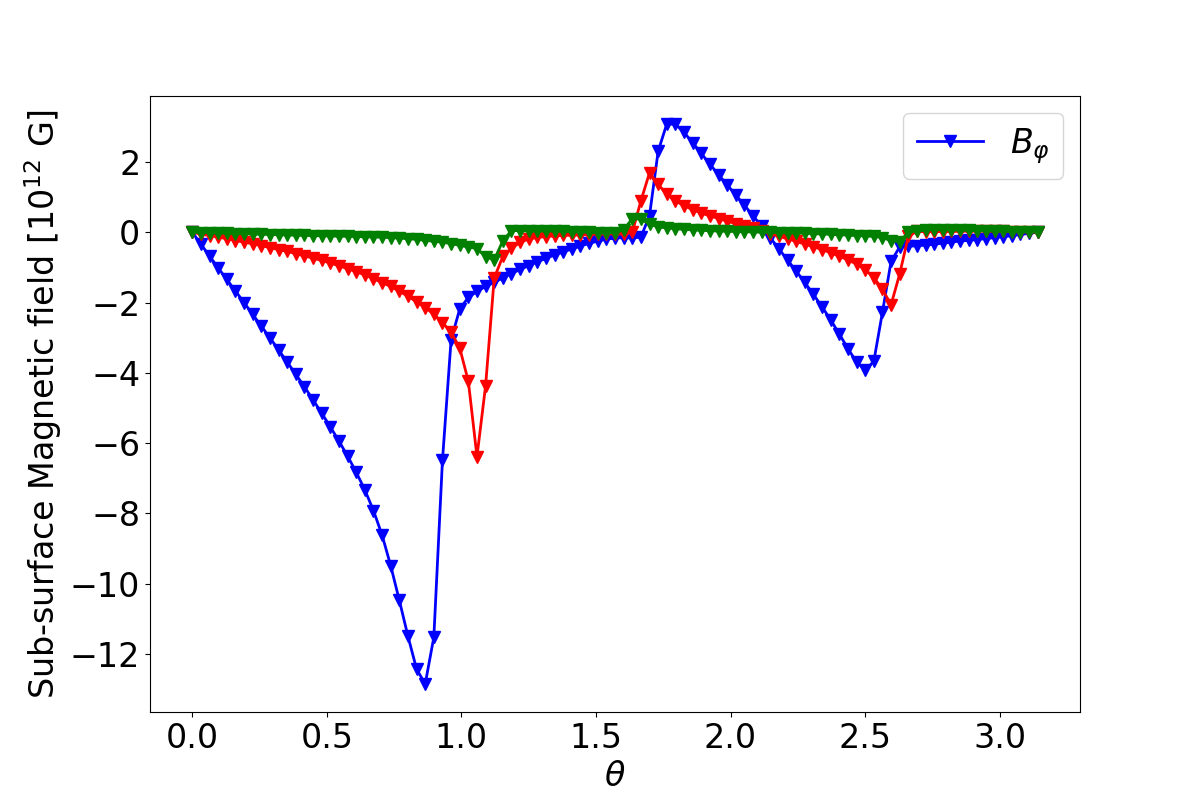}
	\includegraphics[width=\linewidth]{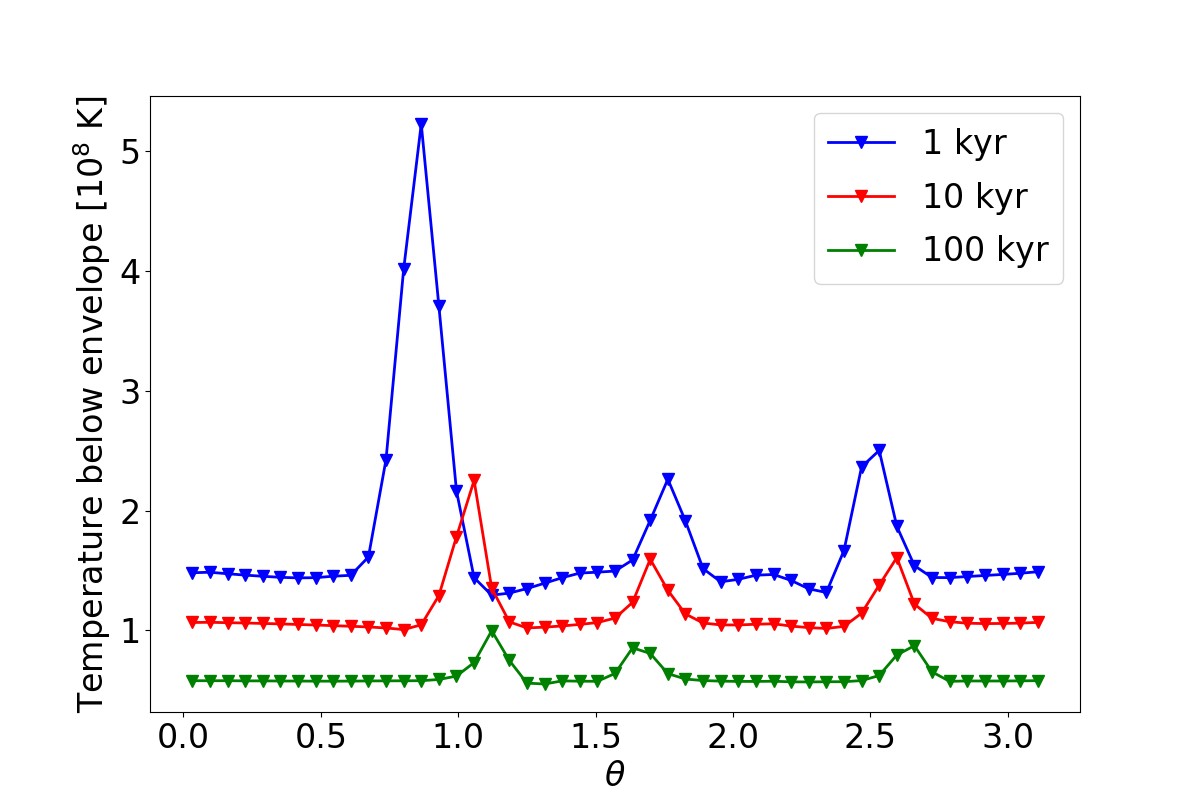}
	\caption{Evolution of the multipolar model {\tt crM} of Fig.~\ref{fig:multipolar_evolution}, showing the meridional profile just below the surface for the poloidal (top) and toroidal (middle) magnetic field components and for the temperature (bottom), at $t=1,10,100$ kyr (blue, red, and green respectively).}
	\label{fig:multipolar_angular}
\end{figure}

As a representative example, we show the evolution of model {\tt crM}, with $N_\theta=49$, $N_r=100$ and the optimal methods discussed above (EULA time advance, $k_{\rm cour}=0.1$, Burgers-like treatment, upwind formulation). In Fig.~\ref{fig:multipolar_evolution} we show the internal distribution of magnetic field (poloidal field lines in white, toroidal field in colors) and temperature, at $t=0,1,10,50$ kyr. Detailed meridional profiles of the magnetic field components and temperature just below the surface are also shown in Fig.~\ref{fig:multipolar_angular}. Throughout the simulations, multiple magnetic poles (where the tangential magnetic field is zero and the field is purely radial) are maintained at $\theta \sim 1.3$ and $ 2.2$, besides the ones imposed by axial symmetry, $\theta=0,\pi$. As one can see, in between the locations of the magnetic poles, strong sheets form, visible as steep profiles in $B_r(\theta)$ and $B_\varphi(\theta)$, and a spike in $B_\theta(\theta)$. These structures last long despite being prone to more dissipation: they are continuously fed and maintained by the Hall dynamics, which compensates the enhanced dissipation.

Note that the capability of the schemes presented here to numerically resolve the formation and evolution of such sharp current sheets is superior to less accurate schemes (for instance, purely centered with no upwind or treatment of the Burgers terms) and to spectral methods, which naturally tend to reconstruct such steep gradients with high multipoles (see for instance the small structures appearing around the discontinuities in Fig. 2 and 3 of \cite{pons07b}).

\begin{figure}[!h]
	\centering
	\includegraphics[width=0.9\linewidth, clip=true]{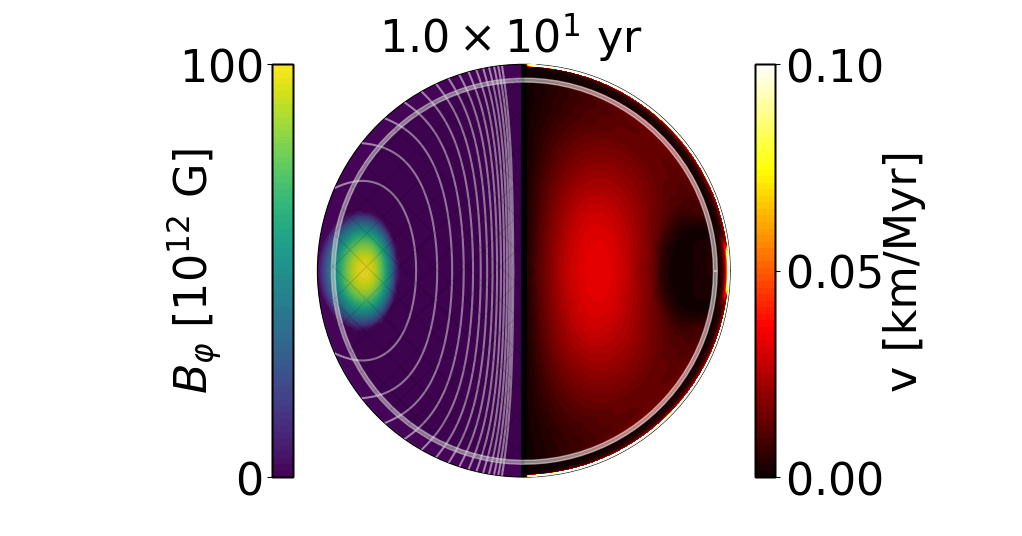}\\
	\includegraphics[width=0.9\linewidth, clip=true]{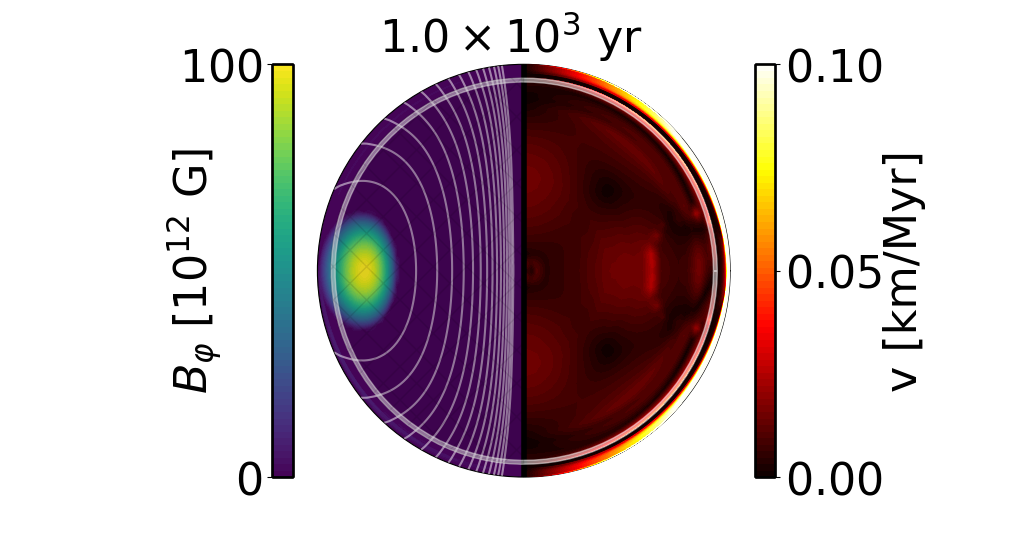}\\
	\includegraphics[width=0.9\linewidth, clip=true]{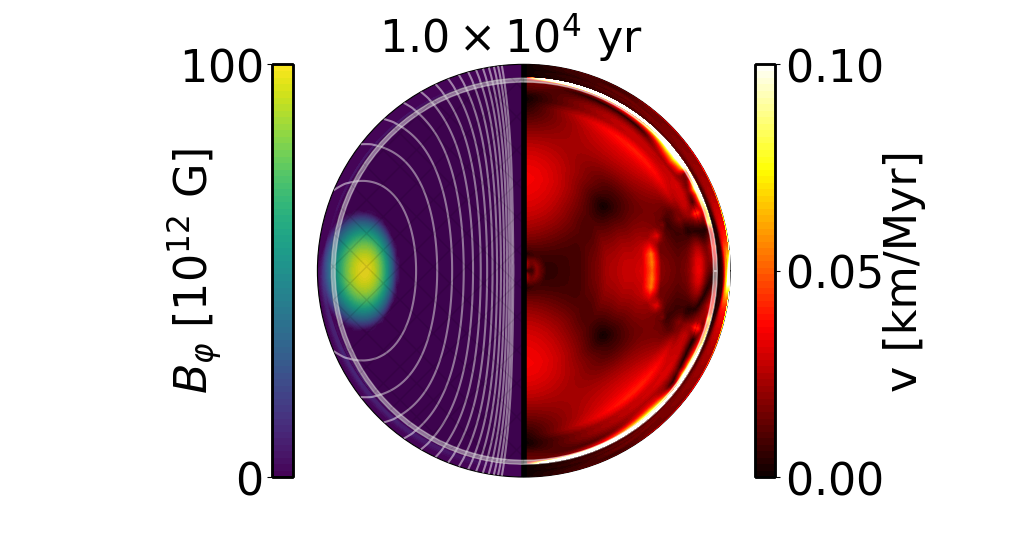}\\
	\includegraphics[width=0.9\linewidth, clip=true]{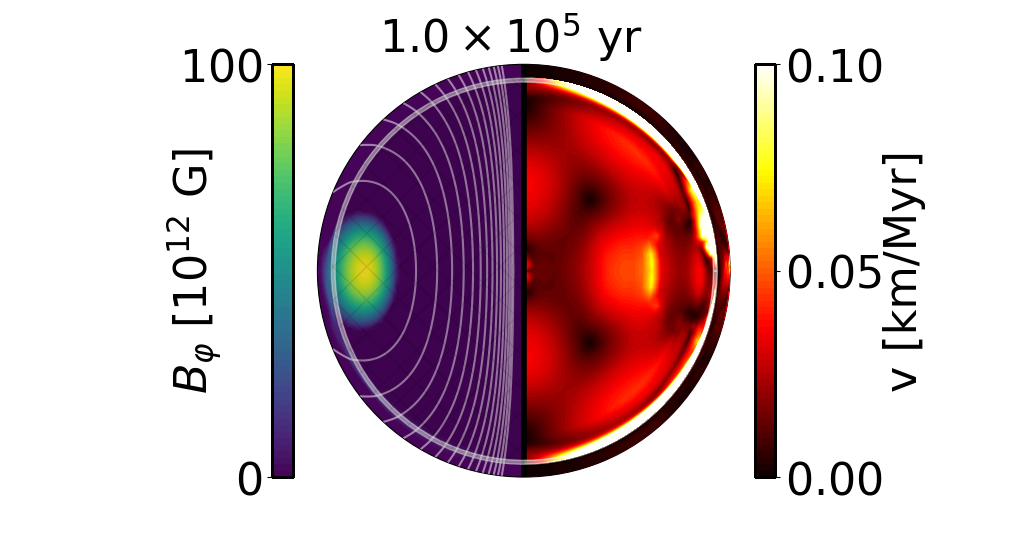}\\
    \caption{Evolution of magnetic field and velocity for the {\tt Core} model at  $t=0.01,1,10,100$ kyr (from top to bottom). The crust-core interface is visible as a solid line. In the left hemisphere we show the meridional projection of the magnetic field lines (white lines) and the toroidal field (colors). In the right hemisphere we show in colors the magnitude of ambipolar velocity (core) and Hall velocities (in the crust, reduced by a factor $10^4$ for visualization clarity). We use $N_r=100$, $N_\theta=49$.}
	\label{fig:ambipolar_evolution}
\end{figure}

\begin{figure}[h!]
	\centering
	\includegraphics[width=\linewidth]{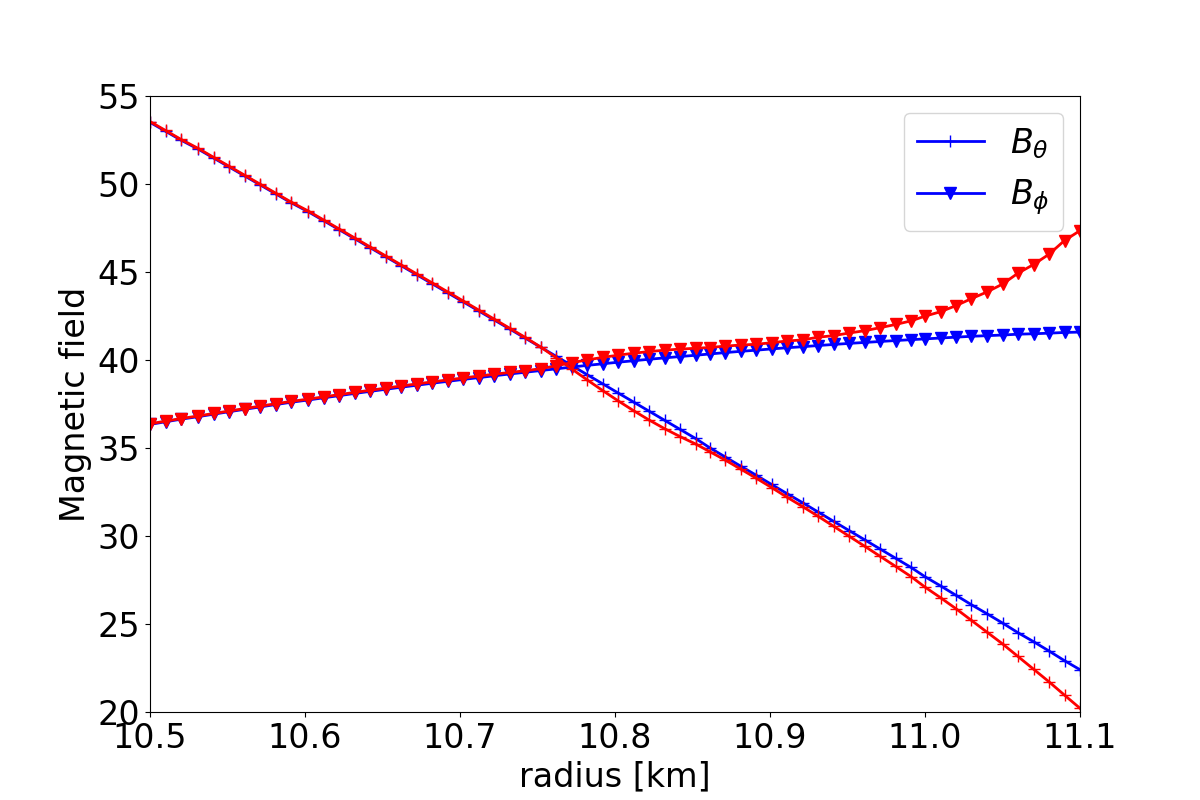}
	\includegraphics[width=\linewidth]{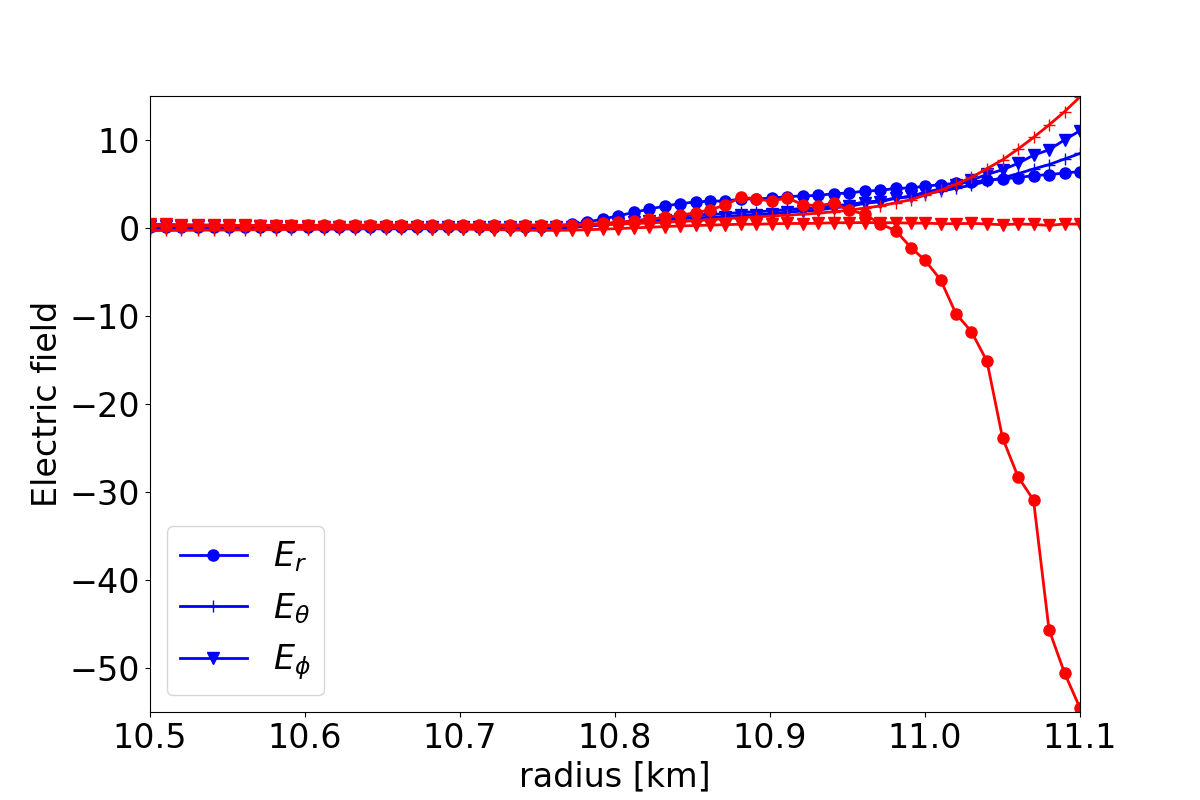}
	\includegraphics[width=\linewidth]{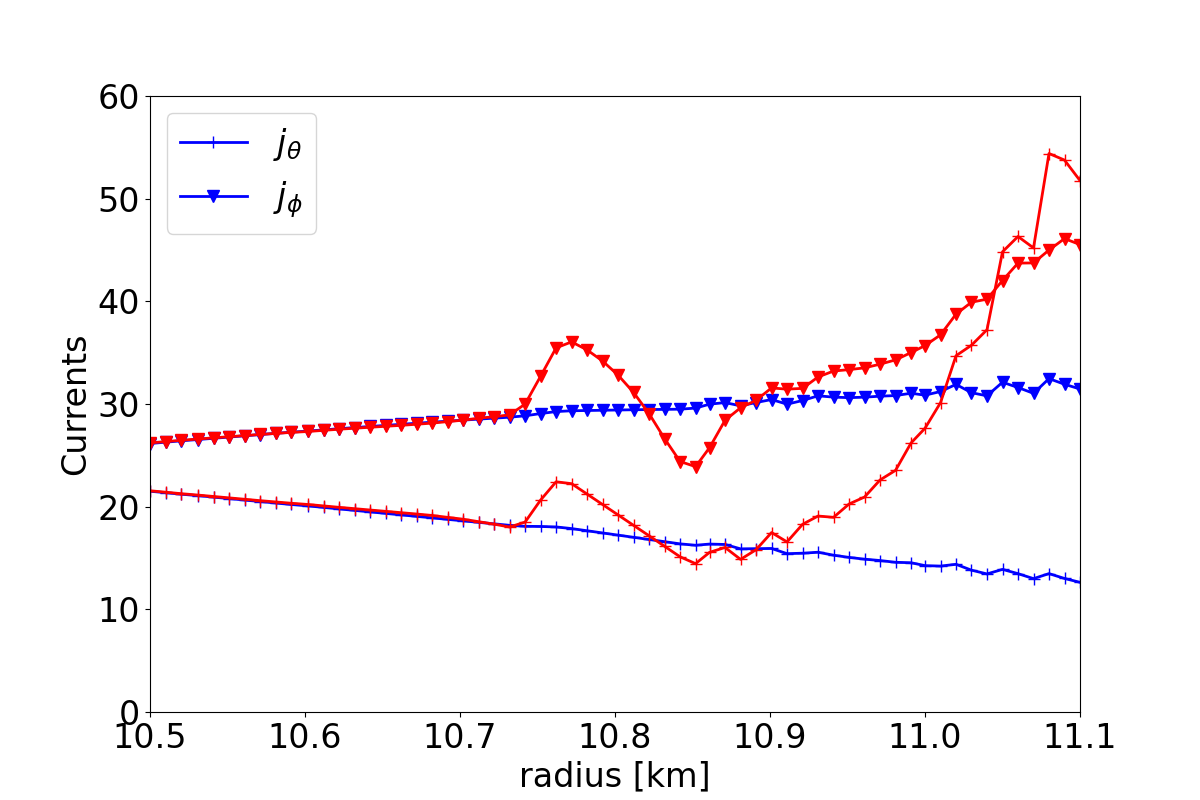}
	\caption{Evolution of the radial profiles at the equator of the three components of the magnetic field (top panel), electric field (middle panel) and electrical currents (bottom panel), for the simulation of Fig.~\ref{fig:ambipolar_evolution}. The components $B_r$ and $j_r$ are omitted as they are very close to zero due to the topology. We show the profile at the transition region between the crust and the core (the interface is located at 10.8 km), at $t=1$ kyr and $t=50$ kyr (blue and red respectively).}
	\label{fig:ambipolar_radial}
\end{figure}

\begin{figure}[!h]
	\centering
    \includegraphics[width=0.75\linewidth, clip=true, trim=120 0 90 0]{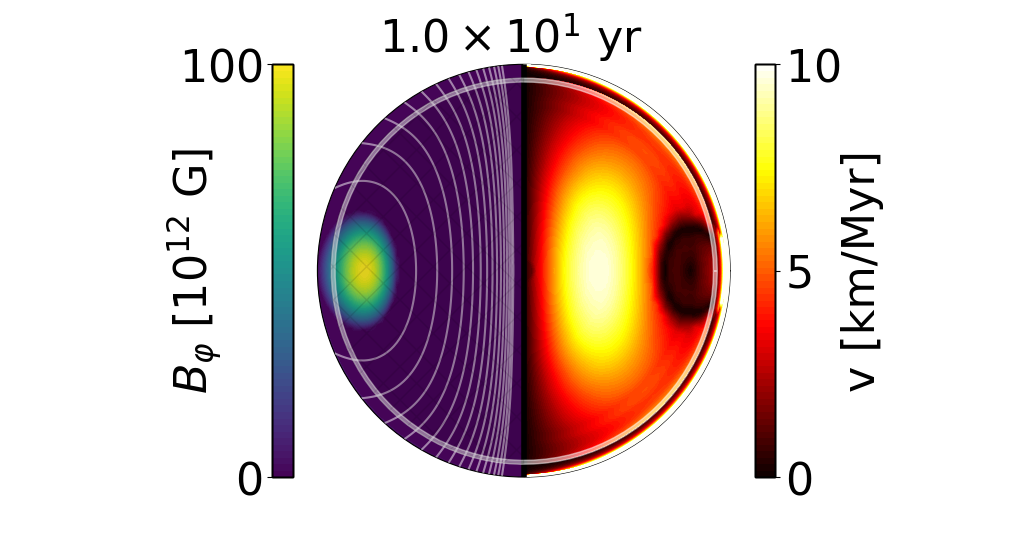}\\
    \includegraphics[width=0.75\linewidth, clip=true, trim=120 0 90 0]{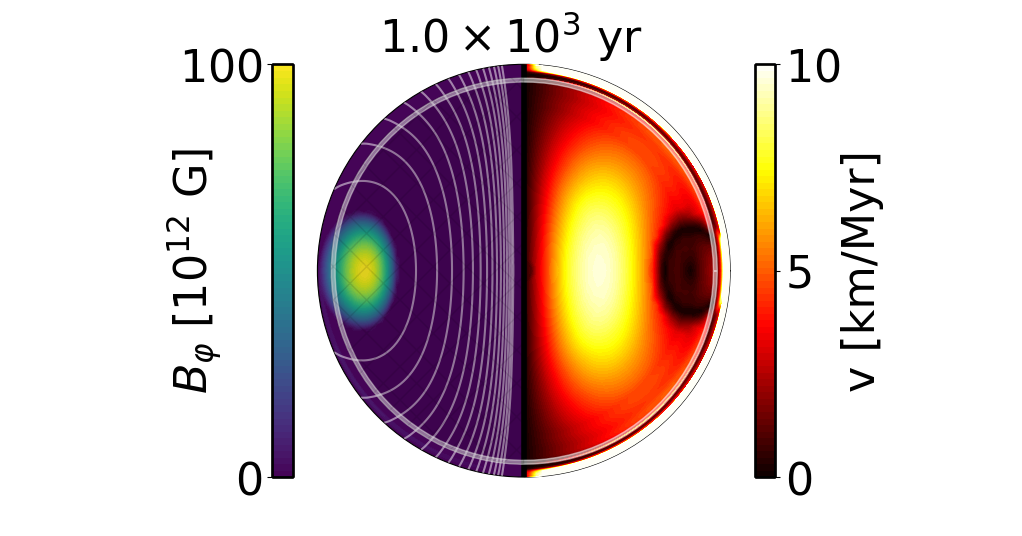}\\
    \includegraphics[width=0.75\linewidth, clip=true, trim=120 0 90 0]{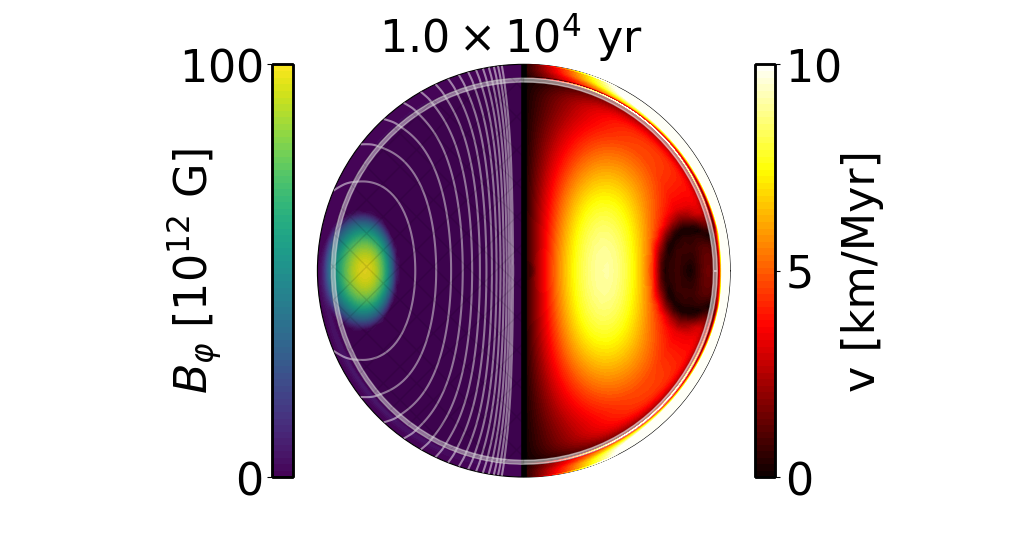}\\
    \includegraphics[width=0.75\linewidth, clip=true, trim=120 0 90 0]{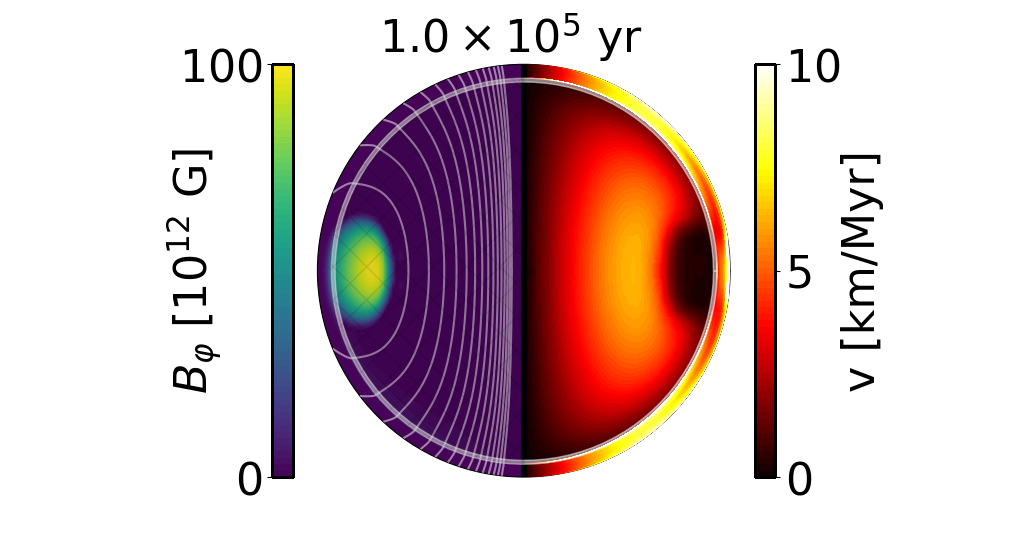}\\
	\caption{Same as Fig.~\ref{fig:ambipolar_evolution} but fixing the temperature $T=10^9$ K, and manually amplifying the ambipolar velocity by a factor 1000 in the simulation. The crustal Hall velocity has been reduced by a factor 10 in the plot, for clarity of visualization.}
	\label{fig:ambipolar_evolution_Tfixed}
\end{figure}

The effects of the anisotropic conduction (induced by the magnetic field) on the temperature are clear by looking at the meridional profiles of the temperature at the outermost crustal layer (bottom panel of Fig.~\ref{fig:multipolar_angular}). Each magnetic pole corresponds to a spike in the temperature, which can be a few times larger than the rest of the star. This kind of behavior is reflected in the surface temperature (for given envelope and emission models) and, eventually, in the spectra and light curves (not treated here, see e.g. \cite{perna13,vigano14,igoshev21,igoshev21b}). Quantifying the effects and interpreting the physical implications is out of the scope of this technical paper and will be dealt with in future works.

\subsection{Core evolution with ambipolar diffusion}
\label{subsec:ambipolar}

Here we show a run with model {\tt Core}, including normal matter (no superfluidity), and the ambipolar diffusion described above. We evolve the temperature and use those values to calculate the ambipolar diffusion coefficients.\footnote{Due to numerical limitations, we enforce a floor value for the temperature entering the calculations of $\tau_{pn}$ and $\lambda$, which control $\vec{v}_a$. Our minimum value, $T^{\rm amb}_{\rm min}=2\times 10^8$ K, is reached after about 3 kyr, which suggests that after this time we might realistically expect higher velocities.} At the same time, we evolve the magnetic fields both in the crust and in the core as explained above. In Fig.~\ref{fig:ambipolar_evolution} we show the evolution of the magnetic field in the left hemisphere. In the right hemisphere we show the velocities: in the core we present $|\vec{v}_a|$, while in the crust the Hall velocity $f_h|\vec{j}|$ (the latter reduced by a factor $10^4$ in order to have the same range).

The main result is that the resulting ambipolar velocity is only a fraction of km/Myr, orders of magnitude below the crustal Hall velocities. Therefore, the magnetic field barely changes over 100 kyr. Looking in more detail at the dynamics, we can see that the crustal field slightly evolves, with a displacement of the toroidal torus, and a bending of the poloidal field lines. The very different timescales of the crust and core evolution naturally tend to create strong currents to support the discontinuity in the tangential magnetic fields.
The pattern of the ambipolar velocities follows those already obtained by \cite{passamonti17a}: at later times, $\Delta \mu$ can in general partially compensate the irrotational part of the Lorentz force. At the same time, velocities tend to be larger due to the smaller reaction rates. As a result, the pattern at late times is more complicated than at early times, with a peak in the velocities close to the crust-core interface.

In Fig.~\ref{fig:ambipolar_radial} we show the radial profiles at the equator of $\vec{B}$, $\vec{E}$ and $\vec{j}$, around the crust-core interface (located in the middle of the range shown). The smooth electrical profiles (see \S~\ref{sec:cc_interface}) are a key element to have a stable run. In the absence of such a transition region, the peaks visible in $j_\theta$ and $j_\phi$ at each side of the interface would be much more pronounced. The calculation of $\vec{v}_a$ depends strongly (via boundary conditions $v_a^r=0$, see \S~\ref{sec:core}) on the Lorentz force at the interface, so that if the latter has strong discontinuities, the numerical evolution becomes unstable.

For the sake of clarity only, we show in Fig.~\ref{fig:ambipolar_evolution_Tfixed} the same {\tt Core} model, but where we have fixed the temperature to $T=10^9$ K and multiplied the ambipolar velocity by a factor 1000. In this case, velocities are artificially higher and the evolution proceeds faster in the core, although still slower than in the crust. Therefore, no visible changes are seen in the core topology. The ambipolar velocity pattern is smoother than in the $T$-evolving case. The higher Ohmic diffusion in the crust (due to the high temperature) causes the Hall dynamics to become less important and weaker discontinuities appear at the crust-core interface.

In general, the simulations performed here show that, since the ambipolar velocities are small and the bulk of currents and magnetic energy is located in the core, their dissipation is very slow. Compared to crust-confined configurations with the same initial dipolar fields, this implies: (i) a lower X-ray luminosity, (ii) a barely evolving dipolar field value, which is closely linked to the rotational evolution and in turn causes (iii) much longer periods (because higher electromagnetic torques are maintained). However, this is directly linked to our simplifying model assumptions. Slow evolution due to intrinsically low velocities is a result of non-superfluid matter, with no direct Urca present. Moreover, the initial configuration of our field has very large scales. Therefore, in reality the evolution could be faster, if (i) superfluidity and superconductivity was included, (ii) the star was massive enough to activate the direct Urca channel, (iii) the magnetic energy was stored in smaller scales, or (iv) one is able to cure the numerical instabilities arising from the ambipolar calculations in presence of low temperatures (high velocities).

\section{Conclusions}
\label{sec:conclusions}

In this work we have published some detailed techniques that are needed to build a robust code for magneto-thermal evolution of neutron stars using finite volumes/finite differences and spherical coordinates.

The lastest version of the axially symmetric magneto-thermal code presented here in detail is faster and more versatile than previous implementations, since it includes alternative numerical methods for finite-volume schemes. The general improvement of the code, after an analysis of the computational bottlenecks, has allowed us to gain a speedup factor of $\sim \times 5-10$ in the overall CPU time (for the same infrastructure and input parameters of the crust-confined models), compared to the version used in e.g. \cite{vigano13}. Such improvements are mostly due to the simplification of existing routines, the use of implicit {\tt Fortran90/95} functions and subroutines, and taking advantage of vectorized operations. We have shown how the local calculations of microphysics represent the main computational bottleneck being, at the same time, a fundamental ingredient for realistic simulations.

The evolution of the induction equation in the crust suffers from a severe timestep constraint inherent to its non-linearity, but it takes only a few percent share of the computational cost. On the other side, it needs special treatment in terms of numerical techniques. From the careful assessment of different discretization methods within a full (not staggered) discrete grid, we conclude that two ingredients are fundamental to resolve the magnetic discontinuities, naturally arising in eMHD: (i) a simple upwind method in the definition of the toroidal electric field (determining the poloidal magnetic field) and (ii) a Burgers-like finite-difference formulation for the Hall part of the toroidal component of the induction equation. Without them, the range of applicability (magnetic field strengths and ages) reduces and the code is only partially applicable to magnetars.

Among the tested time advance methods, EULA is the computationally most convenient one, having its optimal $k_{\rm cour}$ little dependence on the spatial resolution and on the scenario considered. RK4 is slightly slower than EULA for radial resolutions $N_r \lesssim 100$ and hot temperatures (i.e., low magnetization parameter), but is more prone to instabilities for finer resolutions and lower temperatures. The implementation of other methods (EUL, AB4 and the implicit scheme based on pseudospectral methods like in \cite{pons07b}) is much slower and computationally expensive.

Moreover, numerical instabilities in the crust tend to arise due to the Hall effect and strong gradients of $n_e$, especially at early ($\lesssim 1$ kyr) and late ($\gtrsim 100$ kyr) stages. The latter can be at least partially cured by the careful addition of a hyper-resistivity term, which does not change the global solution, and a substantial decrease of the timestep.

Generally speaking, the range of validity of the code, for which instabilities can be totally absent, can be defined as $t\lesssim 10^5$ yr, with initial magnetic fields that, if confined to the crust, can reach up to a few times $10^{14}$ G for the poloidal dipolar component, and a large-scale toroidal field of the same order of magnitude in terms of energy (comparing only its maximum value or the polar surface value of the dipole can be misleading, since what matters is the energy). Simulations with higher initial multipoles and/or higher magnetic fields are possible but more prone to non-negligible numerical instabilities, affecting also the luminosity and possibly disrupting the solution (therefore they should be done with the due attention when drawing conclusions on these results).

We stress that finite-volume/finite-difference methods are able to capture the Hall-driven magnetic discontinuities, which are fundamental to resolve the details of the internal magnetic topology and, as a consequence, of the surface map. Spectral methods, on which a majority of current and past studies are based, cannot offer by construction such capability and the range of reliable applicability is therefore more limited.

An important novelty in the present work is the inclusion of the ambipolar diffusion in the core, using the recipe by \cite{passamonti17b}, consisting of calculating chemical potential deviations. We have smoothly matched the electric fields in the crust and in the core, therefore effectively coupling the evolution of the magnetic field in the two regions, without any (arguably unphysical) sharp current sheets at the interface. The main result is that the timescales for the case considered here (modified Urca process, no superfluidity/superconductivity), the timescales are much longer than the $\sim 1-100$ kyr required to explain magnetars' transient activity and persistent high X-ray luminosities, arising from the dissipation of the currents.

We have also shown how a non-trivial crust-confined topology can be maintained throughout the active life of a magnetar, in agreement with 3D magnetic evolution simulations \cite{gourgouliatos16,igoshev21,igoshev21b}. This, and the tangled magnetic fields produced in recent core-collapse simulations \cite{mosta15} (which should be related to our initial data) reinforce the idea that pure large-scale magnetic fields are likely unrealistic.
Complex topology should be the rule rather than the exception, finding also increasing (albeit indirect) support through observations of old ($\gtrsim 10^5$ yr) neutron stars \cite{tiengo13,borghese15,riley19}.

In general, previous results shown in \cite{vigano13} hold if the same initial crust-confined configurations are used, with minor modifications of luminosity, due to updates of the microphysics and envelope models mainly. The luminosity for the {\tt Core} model considered in this paper is well below the one for the crust-confined models (for a fixed value of $B_{\rm dip}$). This is due to the fact that in the {\tt Core} model the curvature radius of the initial magnetic field lines is about ten times larger and most of the currents circulate in the core: therefore the total currents circulating in the crust are much less than in the crust-confined cases. Moreover, the magnetic field is coupled to the core evolution, which is much slower, at least in the case considered. However, several effects are expected to potentially make the evolution in the core faster: more realistic and complex initial topology, the inclusion of direct Urca processes and the implementation of superconductivity and superfluidity. Future numerical studies will include and study these effects.

The conclusions drawn from this study of methods will be considered in the future 3D extension of the code. Breaking axial symmetry implies that important differences have to be taken into account, among which are the following: (i) if finite differences/finite volumes are used and one coordinate is the radial distance from the center of the star, then one needs to use more than one system of coordinates to avoid the axis singularity of spherical coordinates; (ii) the meridional and azimuthal components are mixed in the poloidal and toroidal components; (iii) the solenoidal constraint and the conservation laws have to be numerically respected considering the full dependence on the three coordinates (in particular, the EULA method and the Laplacian-based hyper-resistivity presented here would introduce a non-zero divergence of $\vec{B}$).

These intrinsic differences imply that the EULA advance and the Burgers-like correction in the discretized induction equation cannot be applied as in 2D. However, all remaining elements are applicable to a 3D code: the logical structure of the code, the microphysics, the cooling scheme (adapted to the 3D grid), and the rest of the magnetic field evolution techniques. Moreover, the stability studies and the computational assessment will be fundamental in guiding the development of a 3D magneto-thermal evolution code.

\subsection*{Acknowledgments}
DV, AGG, CD and VG are supported by the ERC Consolidator Grant ``MAGNESIA" (nr.817661) and acknowledge funding from grants SGR2017-1383 and PGC2018-095512-BI00. JAP acknowledges support by the Generalitat Valenciana (PROMETEO/2019/071), AEI grant PGC2018-095984-B-I00 and the Alexander von Humboldt Stiftung through a Humboldt Research Award. DV acknowledges his Short Term Scientific Mission in Durham (UK) funded by the COST Action PHAROS (CA16214). We acknowledge Nanda Rea for useful comments. The data production, processing and analysis tools have been developed, implemented and operated in collaboration with the Port d'Informaci\'o Cient\'ifica (PIC) data center. PIC is maintained through a consortium of the Institut de F\'isica d'Altes Energies (IFAE) and the Centro de Investigaciones Energ\'eticas, Medioambientales y Tecnol\'ogicas (Ciemat).

\bibliography{biblio}
\bibliographystyle{elsarticle-num}

\appendix

\section{Why not Cartesian coordinates?}\label{app:cartesian}

\begin{figure}[ht!]
	\centering
	\includegraphics[width=0.45\textwidth]{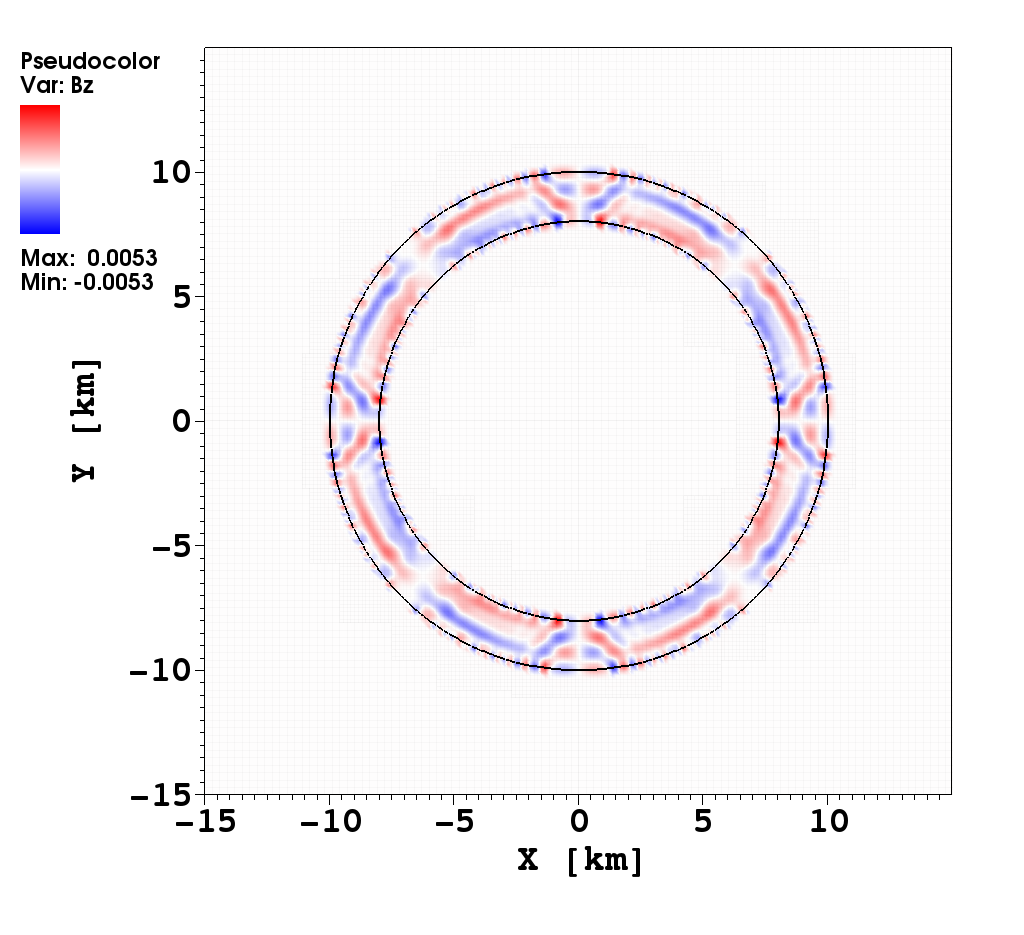}\\
	\includegraphics[width=0.45\textwidth]{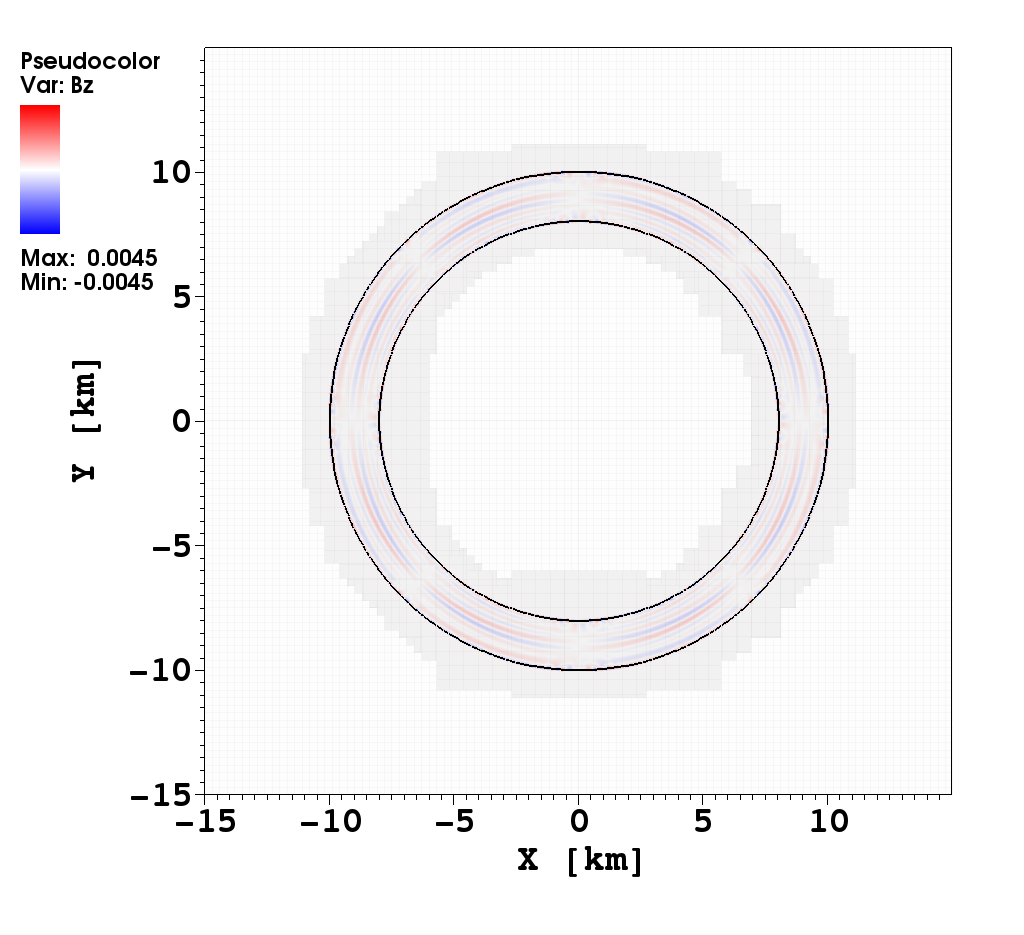}
	\caption{Numerical noise typically arising from the alternative choice of Cartesian discretization of the surface, using the adaptive mesh refinement-based code {\tt Simflowny} \cite{vigano19}. The initial field is purely toroidal, so that no poloidal field should develop, $f_h$ is uniform, with $\eta=0$. The plot shows how a vertical magnetic field in the equatorial plane (corresponding to the poloidal field component $-B_\theta$) develops. We show the same set-up (an extended domain covered by $100^3$ points), refined by a factor 2 (top) and 8 (bottom) in the region covering the crust.}
	\label{fig:cartesian_noise}
\end{figure}

An obvious alternative to a spherical grid is to use Cartesian coordinates, widely used in the MHD community, and tested for the eMHD induction equation for the first time in \cite{vigano19}. They have the advantage that they simplify some geometrical factors in the operators and do not present any singularity on the axis, allowing an easy extension to 3D. However, this choice suffers from two main intrinsic problems: (i) it implies a much higher computational cost, due to the fact that you have to refine all directions even if you want to better resolve the radial gradients only; (ii) the projection of the spherical surface onto the Cartesian grid introduces spurious noise with characteristic patterns, as shown in Fig.~\ref{fig:cartesian_noise}. The noise is partially cured by a computationally costly increase of resolution, as shown by the comparison between the top and bottom panel. However, the noise tends to grow in time (unless it is damped by physical/numerical diffusivity), mixing with the physical small-scale whistler waves naturally arising from the system. The bottom line is that these two drawbacks leave spherical coordinates as the most logical option.

\end{document}